\documentclass[final,3p]{elsarticle}
\usepackage{soul}
 \usepackage{graphics}
 \usepackage{multirow}
 \usepackage{graphicx}
\usepackage{caption}
\usepackage{float}
\usepackage{subcaption}
\usepackage{amsmath}
\usepackage{color}
\usepackage{soul}           
\usepackage{ulem}
\usepackage{tikz}
\usepackage{makecell}
\usepackage{wrapfig}
\usepackage{algorithm}
\usepackage{algpseudocode}
\usetikzlibrary{shapes.geometric, arrows}
\usepackage[numbers]{natbib} 
\tikzstyle{normal} = [rectangle, rounded corners, minimum width=3cm, minimum height=1cm,text centered, draw=black, fill=red!30,text width=10em]

\tikzstyle{robust} = [trapezium, trapezium left angle=70, trapezium right angle=110, minimum width=3cm, minimum height=1cm, text centered, draw=black, fill=blue!30,text width=4em]

\usepackage[T1]{fontenc}
\usepackage{hyperref}
\hypersetup{
colorlinks=true
}
\usepackage{epstopdf}
\usepackage{xcolor}

\usepackage{enumitem}  

\biboptions{sort&compress}

\journal{Measurement}

\begin{document}

\begin{frontmatter}

\title{Robust variance estimators in application  to segmentation of measurement data distorted by impulsive and non-Gaussian noise}

\author[label1]{Justyna Witulska}\corref{cor1}\ead{justyna.witulska@pwr.edu.pl}
\author[label2]{Anna Zaleska } 
\author[label2]{Natalia Kremzer-Osiadacz}
\author[label1]{Agnieszka Wy\l oma\'nska}
\author[label3,label4]{Ireneusz Jabłoński} 

\cortext[cor1]{Corresponding author.}

\address[label1]{Faculty of Pure and Applied Mathematics, Hugo Steinhaus Center, Wroclaw University of Science and Technology, Wyspianskiego 27, 50-370 Wroclaw, Poland}
\address[label2]{Department of Internal Medicine, Pneumonology and Allergology, Wroclaw Medical
University, Poland}
\address[label3]{Brandenburg University of Technology, Cottbus, Germany, (e-mail: jablons@b-tu.de)}
\address[label4]{Fraunhofer Institute for Photonic Microsystems, Cottbus, Germany, (e-mail: ireneusz.jablonski@ipms.fraunhofer.de)}

\begin{abstract}
The paper algorithmizes the problem of regime change point identification for data measured in a system exhibiting impulsive behaviors. This is a fundamental challenge for annotation of measurement data relevant, e.g., for designing data-driven autonomous systems. The contribution consists in the formulation of an offline robust methodology based on the classical approach for structural break detection. The problem of data segmentation is considered in the context of scale change, which physically can be translated into the occurrence of a critical event that reorganizes the system structure. The main advantage of our approach is that it does not require the existence of a variance of the data distribution. The efficiency has been evaluated for simulated data from two distributions and for real-world datasets measured in financial, mechanical, and medical systems. Simulation studies show that in the most challenging case, the error in estimating regime change is 20 times smaller for robust approach compared to the classical one. 

\end{abstract}

\begin{keyword}
robust statistics\sep measurement data segmentation\sep heavy-tailed distribution\sep infinite-variance distribution\sep autonomous systems
\end{keyword}

\end{frontmatter}

\section{Introduction}

Measurement science plays a key role in automating the monitoring of a complex system and management of its resources, which is relevant to modern industry (e.g., Industry 4.0/5.0), medicine (including the 4.0/5.0 medicine concept), business operations, environmental monitoring, realization of the smart city conception, etc. \citep{Macii_2023, Zeb_2024, Branko_2022}.  
Developments in measurement methodologies and instrumentation have revealed that Gaussian processes are not the only mechanism that governs noisy processes, and even the temporal evolution of a complex system can emerge from physical rules far beyond Gaussianity, such as non-Gaussian quantum sensing, non-Gaussian communication, etc.\citep{RAKHUBOVSKY2024, Comunication_non_gauss}. On the other hand, mathematicians have been exploring the abstract space of non-Gaussianity \citep{Nolan2020,shao22}, providing more and more powerful methods and tools for physicists and domain-expert engineers to characterize that class of systems and processes. The signal processing domain has been looking for a robust solution to handle (automatically) a diversity of patterns represented in experimentally measured data at one or multiple sensor nodes, adapting classical transform schemes, including signal filtering, segmentation, decomposition, prediction, etc. \citep{Signal_Processing_General}. Starting from offline and univariate analysis, one can even find algorithms for online exploration of complex data patterns, including filtering and prediction of heavy-tailed components \citep{Provost2023adaptive, SZAREK2023120588}. Given the characteristics of the process (Gaussian/non-Gaussian/mixed, low/significant dynamics, low/high Signal-to-Noise ratio, etc.) and the resources available for measurement action and signal processing (e.g. one/multiple sensor nodes, limited computing power and memory like in edge sensing \citep{NAYAK_2022}) serial operating with collected samples - e.g. filtering, segmentation, prediction - can be inefficient, and in dynamic autonomous systems we should be able to reliably segment raw complex data biased with significant impulse and heavy-tailed distortions.

In this paper we discuss the problem of offline segmentation (estimation of change point) of non-Gaussian data. From a mathematical point of view, this issue can be considered as the problem of "splitting" of random sample (independent identically distributed observations) into two samples of the same probabilistic properties. The main focus is on the case where the separate segments (samples) differ on scales. However, the proposed methodology can be considered as a general problem, i.e. when some characteristics of the data change over time. 
Thus, in our research, we assume that a random sample $X_1,...,X_N$ corresponds to the following model:
\begin{align}
X_i \overset{d}{=} \begin{cases}
     X^{(1)} \sim F(\bar{p}, \gamma_1), \quad \text{for } i < n^*,\\
     X^{(2)} \sim F(\bar{p}, \gamma_2) \quad \text{for } i \geq n^*,
    \end{cases}
    \label{eg:assumption_one_cp}
\end{align}
where $n^*$ is a change point, $F$ represents distribution of random variables $X_1,...,X_N$, $\bar{p}$ is vector of parameters, and $\gamma_1, \gamma_2$ are the scales such that $\gamma_1 \neq \gamma_2$. 

\begin{figure}[h!]
    \centering
    \includegraphics[width=0.7\textwidth]{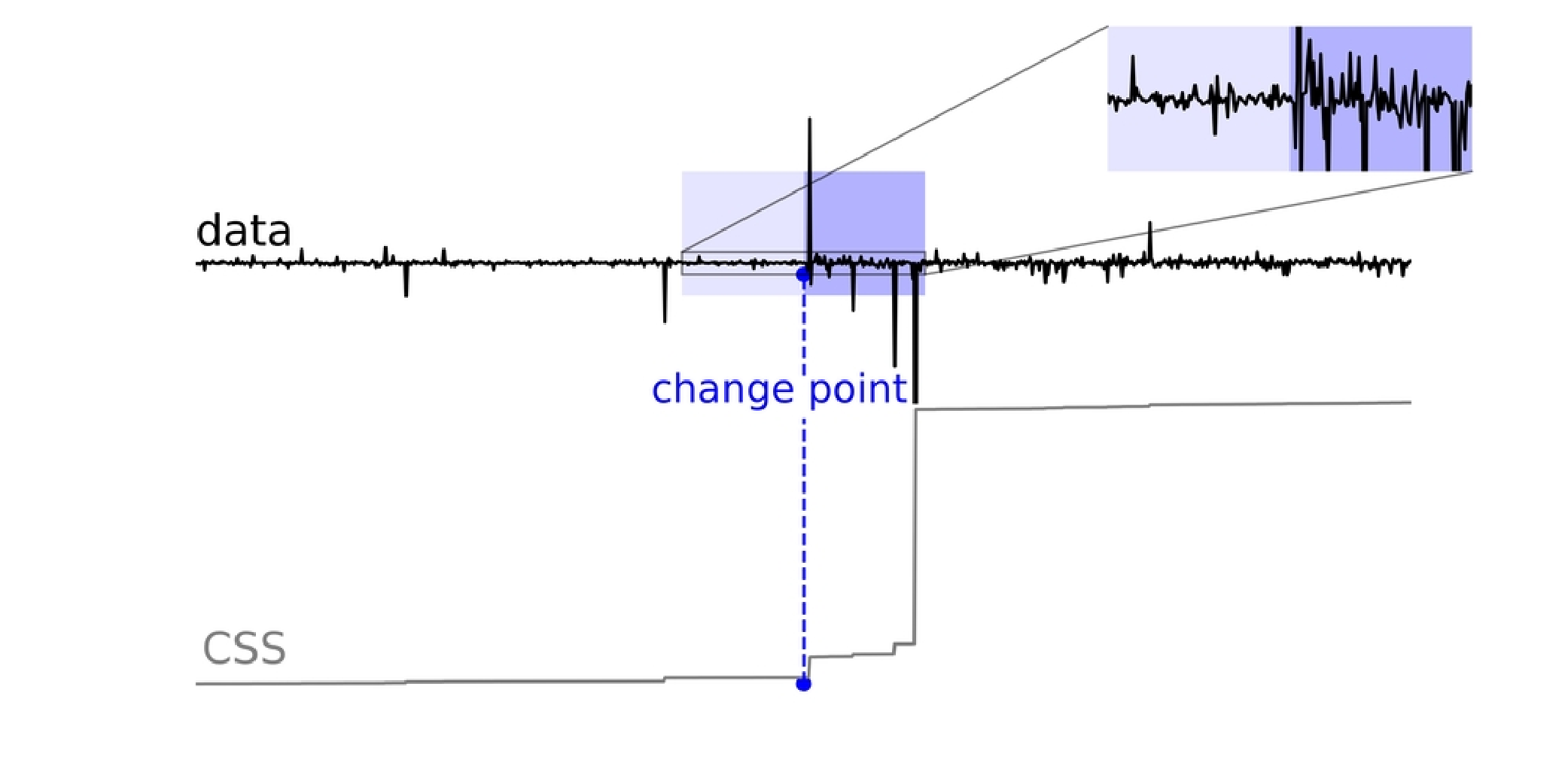}
    \caption{Schematic diagram illustrating the discussed problem of change point detection of heavy-tailed impulsive data.}
    \label{fig:scheme_met}
\end{figure}

As the main focus is on the non-Gaussian data, the problem of segmentation seems to be more challenging in comparison to the classical approaches, where Gaussian (or any finite-variance distribution) is assumed. In the literature, one can find various approaches dedicated to data segmentation. We refer the reader to Section \ref{sota}, where the state of the art is presented. The main goal of our research is to provide a relatively easy, effective, and intuitive method of segmenting non-Gaussian data with a possible infinite-variance distribution. Here, we develop two segmentation procedures based on cumulative sums of squares (CSS) that are designated for data with finite variance, namely Iterative Cumulative Sums of Squares \citep{inclan1994icss} and the quantile method utilizing the ordinary least squares technique \citep{sikora2012regime} considered in the literature as the classical approaches for segmenting samples with changing variance.  The CSS  after a simple transformation can be considered as a statistic that estimates the scale of the distribution of the considered random sample.   However, when the data exhibit impulsive heavy-tailed behavior, the CSS statistic may be an ineffective estimator of the scale (variance) as it is sensitive for large observations occurring in the time series. As a consequence, the specific behavior of CSS statistics resulting from the scale change may be disturbed. In Fig. \ref{fig:scheme_met} we present the schematic diagram that illustrates the considered problem. When the data exhibits impulsive behavior, the classical statistic (here CSS) may not indicate the change point precisely and the algorithms based on it may be inefficient. Thus, we propose robust versions of the classical techniques using robust estimators of scale parameters in the classical algorithms for data segmentation. Despite the proposed methodology is demonstrated for two robust estimators of scale, it can be considered as universal and any other known robust estimator could be applied for the general framework. 

One can list three major advantages of the proposed methodology. First, the proposed method is  nonparametric which means that no assumptions about data distribution are needed. Moreover, the new approach does not require the existence of variance of the considered distribution. Finally, our method is an effective and simple adaptation of well-known techniques with a clear interpretation, and thus, it can be easily adopted in real environments. The efficiency of the proposed approach is verified for simulated data from two different distributions, namely $\alpha-$stable distribution \citep{nolan2012stable} and a mixture Gaussian model \citep{bouguila2020mixture}. Both distributions belong to the class of non-Gaussian distributions; however, they exhibit different probabilistic properties. 

The $\alpha$-stable distribution can be viewed as an extension of the Gaussian distribution ~\citep{stable3}. The class of $\alpha-$stable distributions has found widespread applications in various fields, with common use in financial markets \citep{mand}, physics \citep{phys_new1}, and medicine (specifically heartbeats) \citep{phys35}. The mixture Gaussian model is defined as the sum of the Gaussian distribution with additive outliers defined as uniformly distributed random variables that appear in the Gaussian noise with a given probability. The corresponding random variables have finite variance; however, the probability of extreme values is much higher than in the classical Gaussian distribution. This distribution also found interesting applications, including medicine \citep{cong2014medical}, epidemic spreading \citep{singhal2020modeling}, financial investments, and pricing \citep{lindemann2004probability}.  

The extensive simulation studies are supported by the analysis of real data in three different areas: financial market, mechanical vibrations generated while riding a motorcycle, and acoustic body sounds measured in patients with COVID-19. The simulation study presented and the real data analysis clearly confirm the efficiency and universality of the proposed approach for segmenting the data with non-Gaussian characteristics.

The rest of the paper is organized as follows. In Section \ref{sota} we present state of the art demonstrating different approaches to data segmentation. Section \ref{sec:methodology} introduces the theoretical details of the proposed methodology. In Sections \ref{sec:simulated_data} and Section \ref{sec:real_data}, we compare the results of the proposed approach and the baseline procedures when applied to simulated data and real data sets. Section \ref{sec_Discussion} concludes the paper. All additional remarks and results are presented in the Appendix.

\section{State of the art}\label{sota}

The identification of change point  is a crucial problem considered in many applications, e.g., human activities inspection \citep{li2022automatic, montanez2015inertial}, climate anomalies detection \citep{shi2022changepoint}, quality control \citep{liu2023wind, hawkins2005change}, abnormalities detection in the medical data \citep{sander2020automatic, qureshi2023medical} and danger identification by autonomous vehicles \citep{chen2023edge, feng2020deep}. Detection of anomalies can be helpful to recognize unexpected behavior of the process or examined phenomenon.  The procedure for dividing data into homogeneous segments is called segmentation \citep{Raphael1999360,Cho2015475}. It is usually an important preliminary step of data processing, because the analysis of many stationary segments may be easier than the analysis of one complex signal. The signal segmentation statement could be formulated in a different way for various applications due to the signal characterization, the purpose of research, or a way of data acquisition. For example, some segmentation problems can be solved using online methods (i.e. procedures that are designed to detect changes in real time) or offline methods (\textit{a posteriori} methods i.e., the procedures used to segment data that have been collected in advance) \citep{lavielle1999detection, truong2020selective}. What is important, the process of change may be continuous or the change may be abrupt or gradual \citep{bhaduri2022rough}. The algorithms are created to find changes in a one-dimensional signal or in high-dimensional data (e.g., for image segmentation \citep{jain2023oneformer, wu2024medsegdiff} or multivariate time series segmentation \citep{sikora2017elucidating, matteson2014nonparametric}). The segmentation problem may involve detecting a single change or multiple changes (e.g., binary segmentation and the Pruned Exact Linear Time, described in \citep{killick2012optimal}). Segmentation methods can also be systematized according to their implementation background. These include statistical methods \citep{grzesiek2021divergence, chen2010monitoring}, machine learning methods \citep{li2022automatic, wang2021encoding}, model-based methods \citep{janczura2023machine,  chen2019inference}, kernel-based methods \citep{aminikhanghahi2017survey,  ferrari2023online}, graph-based methods \citep{aminikhanghahi2017survey} etc. Statistical methods do not require a large number of training samples, unlike neural networks. They may be based on theoretical properties of the data distribution (parametric methods) or on empirical estimators of descriptive statistics (nonparametric methods). Machine learning methods include all algorithms that improve their effectiveness as the number of training samples increases.

The change can be exhibited by various factors, e.g., mean \citep{hawkins2005statistical, mean6, mean7, li2021adversarially}, variance (/scale) \citep{inclan1994icss, hawkins2005change, sikora2012regime}, heavy-tailed index \citep{jin2018testing}, other distribution parameters or model coefficients \citep{chen2019inference}. It is worth noting that there exists many methods for scale change point detection for data with finite variance \citep{inclan1994icss, hawkins2005change, chapman2020nonparametric}.  

However, these methods do not work when we have a change in scale for heavy-tailed data (especially data with infinite variance). In this research, we focus on developing \textit{a posteriori} methods for detecting scale change points in one-dimensional data. 

The subject of the research is to introduce the methodology  for segmentation of non-Gaussian data, including data with infinite variance. There exists many studies that propose segmentation methods for heavy-tailed data (especially with infinite variance) in the case of changes in the location \citep{li2021adversarially, sankararaman2023online, bazarova2015change}, persistence \citep{chen2012monitoring, chen2010monitoring},  index responsible for the heavy-tailed behavior \citep{jin2018testing}, $\alpha$ parameter (for $\alpha-$stable distribution) \citep{shi2008consistency}. However, there is a clear methodological gap in research related to scale change point detection for data with infinite variance. Because of assumptions, methods for scale change point detection in Gaussian data are ineffective for heavy-tailed data. In practice, long-tail data appears in many areas (e.g. medicine, physical, technical, finance). Therefore, it is necessary to develop procedures that allow data segmentation for such cases.

\section{Methodology}
\label{sec:methodology}
In this section, a novel methodology for data segmentation is proposed. The general idea is to use robust statistics estimating the scale (or variance in finite-variance case) for tuning baseline procedures for change point detection. As mentioned, the selected reference methods for change point detection utilize the CSS statistic which is sensitive for large observations. The application of robust statistics to estimate the scale (instead of classical CSS) seems to be the natural choice here. First, we recall two selected procedures for regime change detection known from the literature. Next, we present two robust estimators of the scale parameter that are further used in robust algorithms. Finally, we briefly discuss how to enhance the proposed approaches to be more efficient for real data. The source code was shared at ResearchGate (DOI: 10.13140/RG.2.2.16332.42889).

\subsection{Baseline algorithms for regime change point detection}
\label{sec:change_point_baseline}
\begin{itemize}
\item \textbf{The Iterative Cumulative Sum of Squares} (ICSS) \\
The ICSS is a parametric method introduced in \citep{inclan1994icss}. The procedure requires a specific form of input data. In such an approach we assume that $X_1, X_2, ..., X_N$ are independent random variables such that $X_n\sim N(0,\sigma_n^2)$ for each $n=1,2,...,N$. The ICSS method is based on a normalized and centered version of CSS defined as follows:
\begin{align}
    S_n = \frac{C_n}{C_N} - \frac{n}{N}, \quad n = 1,2,...,N \text{, \ and \ \ } S_0 = 0,
    \label{eq:icss}
\end{align}
where $C_n$ is CSS statistic defined as:
\begin{align}
    C_n = \sum_{i=1}^{n} X_i^2 \text{, \ \ \ } n = 1,2,...,N.
    \label{eq:cumsum}
\end{align} 
\noindent
The estimated change point obtained by ICSS algorithm is defined as:  
\begin{align}n^* = arg_{n \in \{2,3, ..., N-1\}} max|S_n|.\label{estimator}\end{align}

The assumption of a Gaussian distribution of the random sample is required to identify the asymptotic distribution of $\sqrt{N/2}S_n$. In \citep{inclan1994icss} it is shown that this statistic behaves like the Brownian bridge for large samples. For more details about the Brownian Bridge, see \citep{chow2009brownian}. This fact is utilized in designing a statistical test to determine the significance of the estimated change point. However, the simulation studies indicate that the segmentation method based on the ICSS algorithm is also effective for non-Gaussian data under the assumption that it originates from a finite-variance distribution.
\item \textbf{The quantile method}\\
The quantile method proposed in \citep{sikora2012regime} also utilizes the CSS statistic to identify the change point. However, it does not require the assumption of a Gaussian distribution of the data, and the distribution of test statistics is calculated by Monte Carlo simulations. The only requirement is that the random sample $X_1,...,X_N$ constitutes a sequence of independent random variables from the finite-variance distribution. The statistic used to detect the regime change point is defined as follows:
\begin{align}
        S_n =  \sum_{j=1}^{n}(C_j - (\alpha_{1,n} + \beta_{1,n}\cdot j))^2 + \sum_{j=n+1}^{N}(C_j - (\alpha_{2,n} + \beta_{2,n}\cdot j))^2,
        \label{bkk}
    \end{align}
   where $C_j$, $j=1,2,...,N$ is a statistic defined in \eqref{eq:cumsum}. In the above definition, the parameters $\alpha_{1,n}, \beta_{1,n}$, $\alpha_{2,n}, \beta_{2,n}$ are linear regression coefficients calculated by the ordinary least squares (OLS) method. For more details on OLS, see \citep{gross2003linear}. The first two parameters, i.e. ($\alpha_{1,n}, \beta_{1,n}$) refer to the first $n$ observations and the next two ($\alpha_{2,n}, \beta_{2,n}$) are estimated for the last $N-n$ values. 
    The estimated change point is defined as the value $n$ (from the set $\{2,3,...,N\}$) for which the statistic $S_n$ takes the minimum.  For more details,  see \citep{sikora2012regime, ww2022identification}.

\end{itemize}

\subsection{Robust estimators of scale parameter}
\label{sec:robust_estimators}
In this section, we recall two well-known statistics that serve as robust estimators of the scale parameter (in particular scale) when a random sample is drawn from a non-Gaussian heavy-tailed distribution. Both of these statistics have been examined from a theoretical standpoint and have found applications in various contexts. 

Our initial focus is on demonstrating the connection between the CSS statistic defined in Eq.  (\ref{eq:cumsum}) and the corresponding sample variance. This connection serves as the starting point for our methodology, wherein the classical estimator of variance is replaced by its robust counterparts.

Let $X_1, X_2, ..., X_N$ be a random sample and $n = 1, ..., N$. Let us denote $\hat{\mu}_n=\frac{1}{n}\sum_{i=1}^nX_i$ as a sample mean calculated for the random sample $X_1,X_2,...,X_n$ and $\hat{\sigma}_n^2=\frac{1}{n-1}\sum_{i=1}^n(X_i-\hat{\mu}_n)^2$ as its sample variance. One can easily show that the following holds:
\begin{align*}
   (n-1)\cdot \hat{\sigma}_n^2 = \sum_{i=1}^{n}(X_i - \hat{\mu}_n)^2 = \sum_{i=1}^{n} X_i^2 - 2\hat{\mu}_n \sum_{i=1}^{n} X_i + n \cdot \hat{\mu}_n^2 = C_n - 2\hat{\mu}_n \sum_{i=1}^{n} X_i + n \cdot \hat{\mu}_n^2={C_n-n\hat{\mu}_n^2}. 
\end{align*}

\noindent
Thus, we obtain that CSS statistic defined in Eq. (\ref{eq:cumsum}) can be represented as:
\begin{align}\label{podstawa}
C_n=n\left(\hat{\sigma}_n^2+\hat{\mu}_n^2\right) {- \hat{\sigma}^2_{n}}.
\end{align}
 
As can be observed, the CSS statistic can be derived as a linear combination of the sample variance of $X_1, X_2, ..., X_n$ and its sample mean. However, as mentioned, the sample variance is considered to be statistically sensitive for large observations, resulting from the fact that the data come from a heavy-tailed distribution. Thus, in our methodology to calculate the robust version of CSS, in Eq. (\ref{podstawa}) we replace the sample variance $\hat{\sigma}_n^2$ by the robust estimator of variance  $\hat{\sigma}_{r,n}^2$. Additionally, in some extreme cases even the sample mean $\hat{\mu}_n$ may be influenced by large observations (e.g., in situations when theoretical mean does not exist); thus in this case we replace it by a robust estimator of the theoretical mean, namely the sample median $\hat{\mu}_{r,n}$. Finally, the robust version of the CSS statistic considered in this paper takes the form:
 \begin{align}
    C_{r, n} = n \left(\hat{\sigma}^2_{r,n} + \hat{\mu}_{r,n}^2\right) {- \hat{\sigma}^2_{r,n}}.
    \label{eq:robust_Cj}
\end{align}
In Eq. (\ref{eq:robust_Cj}) $\hat{\mu}_{r,n}$ is a sample median of $X_1,X_2,...,X_n$ while $\hat{\sigma}^2_{r,n}$ is {an arbitrary} robust estimator of variance. In the literature, many robust estimators of the variance can be found.  However, in this paper we apply two of them. They are further denoted as $\hat{\sigma}^2_{r1,n}$ and $\hat{\sigma}^2_{r2,n}$. The appropriate definitions are given below. Note that if we calculate them only for random variables $X_1,X_2,...,X_n$, where $n<N$, then the appropriate estimators are denoted $\hat{\sigma}^2_{r1,n}$ and $\hat{\sigma}^2_{r2,n}$. However, if they are calculated for the entire random sample, that is, for random variables $X_1,X_2,...,X_N$, then for simplicity we take the notation $\hat{\sigma}^2_{r1}$ and $\hat{\sigma}^2_{r2}$, respectively. The same nomenclature is taken in case of sample median statistic. The statistic $C_{r,n}$ that uses $\hat{\sigma}^2_{r1,n}$ as the robust estimator of variance is denoted by $C_{r1,n}$. Similarly $C_{r2,n}$ is the $C_{r,n}$ that utilizes $\hat{\sigma}^2_{r2,n}$.
 
\begin{itemize}
    \item \textbf{The biweight A estimator} (the biweight midvariance; BMID) \citep{lax1985robust, wilcox1993comparing} \\
    Let $ X_1,...,X_N$ be a random sample. The biweight midvariance is defined as follows:
\begin{align}
    \hat{\sigma}^2_{r1}  = N \ \frac{\sum_{|U_i| < 1} \
    (X_i - \hat{\mu}_r)^2 (1 - U_i^2)^4} {(\sum_{|U_i| < 1} \
    (1 - U_i^2) (1 - 5U_i^2))^2},
    \label{eq:BMID}
\end{align}
where $U_i = \frac{(X_i -\hat{\mu}_r )}{c \cdot MAD}$, $c$ is some constant and $MAD$ is a median of a sample $|X_1-\hat{\mu}_{r}|,|X_2-\hat{\mu}_{r}|,...,|X_N-\hat{\mu}_{r}|$. 
{The authors of \citep{lax1985robust} pointed out that the effectiveness of BMID (and its other versions) stands out among other known robust scale estimators. Its effectiveness is confirmed by Monte Carlo simulations conducted for various distributions (also for heavy-tailed data). In the article \citep{wilcox1993comparing} it is mentioned the possibility of using BMID to construct the statistical tests of equality of scales for independent data.

\item The \textbf{sample quantile conditional variance}  (QCV) \citep{pitera2022goodness} for given constants $a$, $b$ (such that $0<a<b<1$) is defined as follows:

\begin{align}
   \hat{\sigma}^2_{r2} = \frac{1}{[Nb]-[Na]}\sum_{i=[Na]+1}^{[Nb]}\left(X_{(i)} - \hat{\mu}(a, b) \right)^2, 
    \label{eq:QCV}
\end{align}
where $X_1,...X_N$ is a random sample, $X_{(i)}$ is the $i$th order statistic of the sample and $\hat{\mu}_X(a, b)$ is the conditional sample mean defined as follows:
\begin{align*}
\hat{\mu}(a, b) = \frac{1}{[Nb]-[Na]}\sum_{i=[Na]+1}^{[Nb]}X_{(i)}.
\end{align*}
The QCV statistic was extensively discussed in \citep{pitera2022goodness}, where its main probabilistic properties were shown for the general class of heavy-tailed distributions (e.g., $\alpha-$ stable distributions). It was also used in the testing and estimation problem for this class of distributions; see \citep{paczek1,paczek2}. }
\end{itemize}

\subsection{Enhancement of the robust regime change point detection algorithms}

In this part, we briefly discuss how to enhance the robust versions of the algorithms for change point detection. Both segmentation procedures require correction due to the asymmetry of the robust CSS statistics. In general, for data with one change point, the statistic $C_{r, n}$ can be a convex or concave function of $n$.

\begin{figure}[h!]
    \centering
\includegraphics[width=0.8\textwidth]{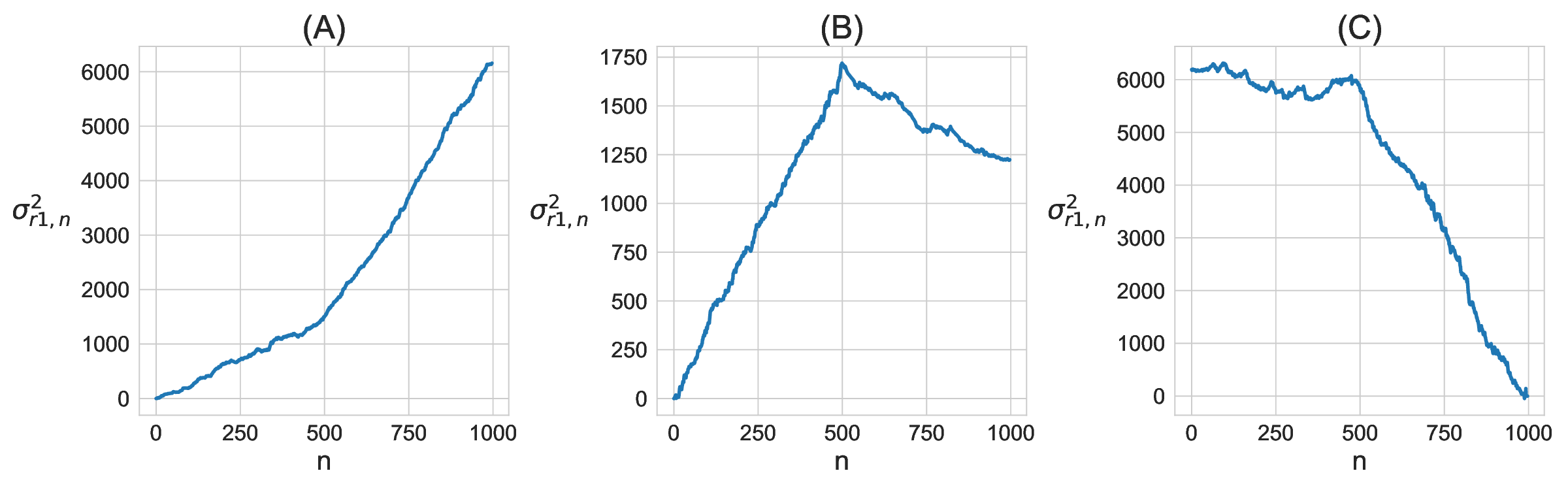}
    \caption{The possible scenarios of $C_{r1, n}$ statistic  behavior for example random sample from symmetric $\alpha$-stable distribution (see the next section for more details) with one change point. Panel A: scale in the first segment equals 1, and in the second segment the scale is equal to 2; B: scale in the first segment equals to 1, and in the second segment the scale is equal to 0.25, C: scale in the first segment equals to 1, and in the second segment the scale is equal to 2; the statistics $\hat{\sigma}_{r1, n}$ for panel C are calculated for reverse data: $X_{N},...,X_{1}$.}
    \label{fig:ex_convexity}
\end{figure}

The illustrative example of the statistic $C_{r, n}$ (taken as $C_{r1, n}$) for both cases (i.e. when it is convex and concave function) is visible in Fig. \ref{fig:ex_convexity} (panels A and B). Both algorithms for change point detection require the concave robust CSS (see panel A). However, they can also be applied if this function is convex. In this case,  we calculate the statistic $C_{r, n}$ for the reverse sample, i.e., for the sample $X_n,...,X_{1}$. One can verify automatically whether $C_{r, n}$ statistic is a concave or convex function of $n$. The simple method is to fit the linear function to the points ($2$,  $C_{r,2}$) and ($N-1$,  $C_{r,N-1}$) and to verify whether $C_{r, n}$ is above the fitted line (concave function) or below it (convex function). 

What is noteworthy, the $C_{r, n}$ statistic calculated for reverse data can resemble a function presented in Fig. \ref{fig:ex_convexity}C (i.e., for initial indices $n$, the $C_{r, n}$ is approximately constant).  If this situation occurs, it is recommended to calculate $C_{r, n}$ for sample $X_1,...,X_n$. 
We can check if initial indices $n$ the $C_{r,n}$ is approximately a constant using a simple procedure:
\begin{itemize}
    \item calculate the mean of  $C_{r,n}$  obtained  for first values of $n$. In our case, we take $n=2,...,7$. The mean we denote as $\omega$,
    \item calculate $\varepsilon = \frac{\#\{n \in \{2,...,N-1\}: \ C_{r,n} > \omega\}}{N-2}$,
    \item if $\varepsilon$ is low (in our case lower than 0.05), then we conclude that for initial $n$ (in our case $n=2,...,7$) the statistic behaves like a constant (case from Fig. 1C).
\end{itemize}

\noindent
To conclude, depending on the input data, the robust statistic $C_{r, n}$ can be calculated for data in the original order or reversed (the first observation as the last and the last as the first).

\section{Efficiency of the considered  robust algorithms for simulated data}
\label{sec:simulated_data}
In this section, we assess the introduced robust procedures for change point detection using simulated data. Analyzes are conducted to validate the effectiveness of individual methods by examining the distance between the detected change points and the true (assumed) one.
 
The correctness of the enhanced segmentation methods is verified via boxplots analysis. We compare baseline methods and their enhanced versions (for two $\hat{\sigma}^2_{r,n}$ statistics). Visual inspection allows us to verify the effectiveness of the introduced solution in the problem of segmentation of heavy-tailed or spiky data. Here, we consider two types of distribution, namely, symmetric $\alpha-$ stable distribution and mixture Gaussian model. The first distribution belongs to the heavy-tailed class, while the data corresponding to the second model exhibit impulsive (spiky) behavior.  In the analysis, we take the following notation:
\begin{itemize}
    \item \textit{ICSS [BMID]} as a robust version of ICSS that uses $C_{r1, n}$, 
    \item \textit{ICSS [QCV]} as a robust version of ICSS that uses $C_{r2, n}$. We take $a=b=0.1$.
    \item \textit{OLS [BMID]} as a robust version of quantile method that uses $C_{r1, n}$, 
    \item \textit{OLS [QCV]} as a robust version of quantile method that uses $C_{r2, n}$. We take $a=b=0.1$.
\end{itemize}

\subsection{Symmetric $\alpha$-stable distribution}
\label{sec:alpha_stable_distr}
\noindent
We recall that the random variable $X$ is symmetric $\alpha-$stable distributed if its characteristic function takes the form \citep{nolan2012stable}: 
\begin{align}
\varphi_X(t) = E[exp(itX)] =  exp(-\gamma^{\alpha}|t|^{\alpha}),
\end{align}
where $0<\alpha\leq 2$ is the stability index, and $\gamma>0$ is the scale parameter.  In this paper, the symmetric $\alpha-$stable distributed random variable we denote as  $\mathcal{S}_{\alpha}(\gamma)$. Let us recall, when $\alpha=2$, the random variable $X$ is Gaussian distributed. For other values of $\alpha$, the $\mathcal{S}_{\alpha}(\gamma)$ random variables have infinite variance. For $\alpha \leq 1$ also the expected value (theoretical mean) does not exist.

In the conducted simulation study we assume that the random sample $X_1,...,X_N$ corresponds to the following model:
\begin{align}
X_i \overset{d}{=} \begin{cases}
     X^{(1)} \sim \mathcal{S}_{\alpha}(\gamma=1), \quad \text{for } i < n^*,\\
     X^{(2)} \sim \mathcal{S}_{\alpha}(\gamma=\gamma_2) \quad \text{for } i \geq n^*,
    \end{cases}
    \label{eq:formula_stable}
\end{align}
where $i=1,2,...N$, $n^*$ is a change point and $\gamma_2$ is the scale parameter corresponding to the second segment. We assume $\gamma_2 \neq 1$. 

\begin{figure}[h!]
    \centering
    \includegraphics[width=0.9\textwidth]{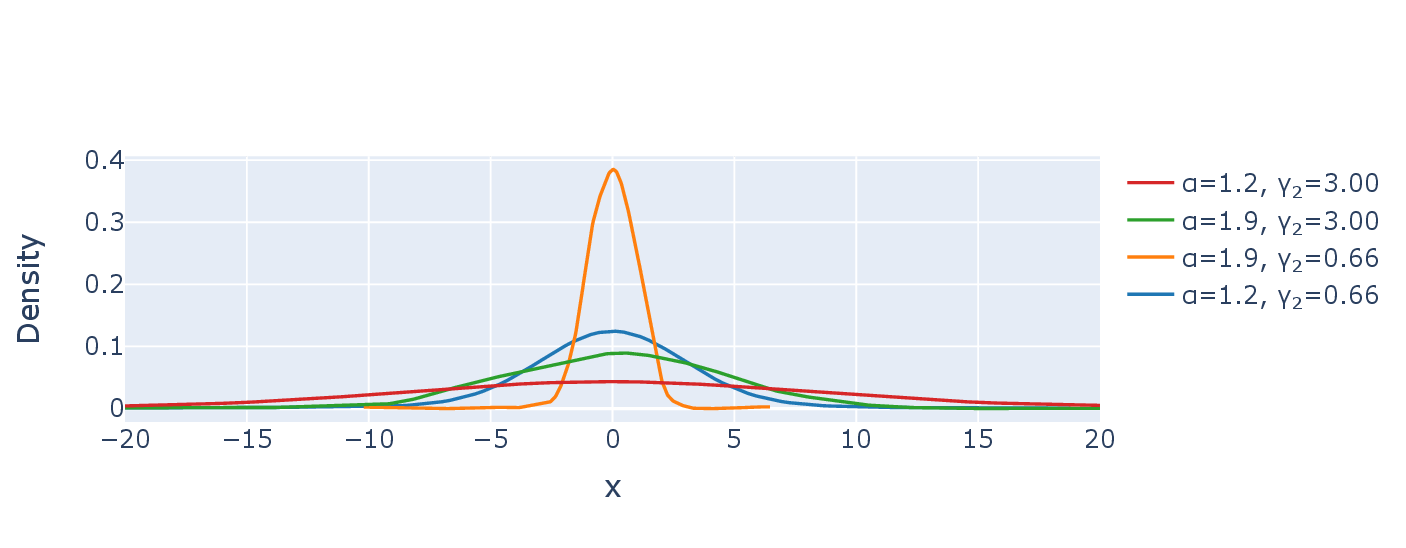}
    \caption{The p.d.f.s for the symmetric $\alpha-$stable distribution (second segment) for different values of the parameters.}
    \label{fig:ex_distr_sbl}
\end{figure}

The probability density functions (p.d.f.) of $\mathcal{S}_{\alpha}(\gamma=\gamma_2)$ distribution for exemplary values of the parameters (corresponding to the second segment from model (\ref{eq:formula_stable})) are presented in Fig. \ref{fig:ex_distr_sbl}. For comparison, we consider four sets of parameters, namely $(\alpha, \gamma_2)$: $(1.2, 0.66)$, $(1.2, 3.00)$, $(1.9, 0.66)$, and $(1.9, 3.00)$.

For selected $\alpha$ we can observe that for smaller $\gamma_2$ (that is, $\gamma_2=0.66$), the values are more concentrated around zero than for $\gamma_2=3$. Importantly, for $\alpha=1.2$ more visible heavy-tailed behavior may be observed than for $\alpha =1.9$. It means that for lower values of $\alpha$ the extreme values are more probable than for higher values of $\alpha$'s. However, even for distributions with $\alpha=1.9$ we can observe that the distribution tails are heavier than in the Gaussian case. 
In the Appendix (Fig. \ref{fig:trajectory_alpha_12_gamma2_066} -- Fig. \ref{fig:trajectory_alpha_19_gamma2_3}) we present the exemplary trajectories corresponding to the model (\ref{eq:formula_stable}) with parameters in the second segment used as in Fig. \ref{fig:ex_distr_sbl}. Here we assume that $N=1000$ and the change point is located in the middle of the data.

\begin{figure}[h!]
    \centering
    \includegraphics[width=0.95\textwidth]{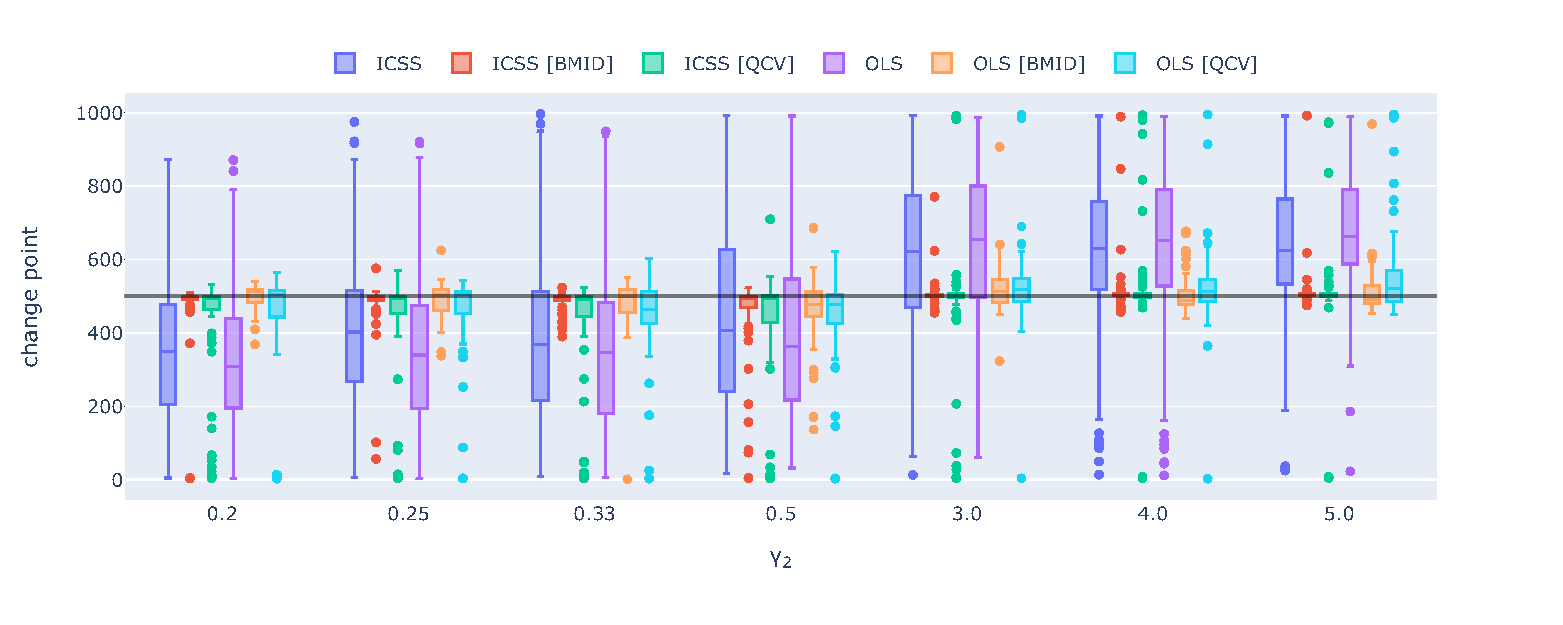}
    \caption{Detected change points obtained using selected procedures for sample described by model \eqref{eq:formula_stable} for $N=1000$. Distribution of the data: symmetric $\alpha-$stable distribution with \uline{$\alpha=1.1$}. True change point is marked via grey line.}
    \label{fig:sim_as_alpha11}
\end{figure}

\begin{figure}[h!]
    \centering
    \includegraphics[width=0.95\textwidth]{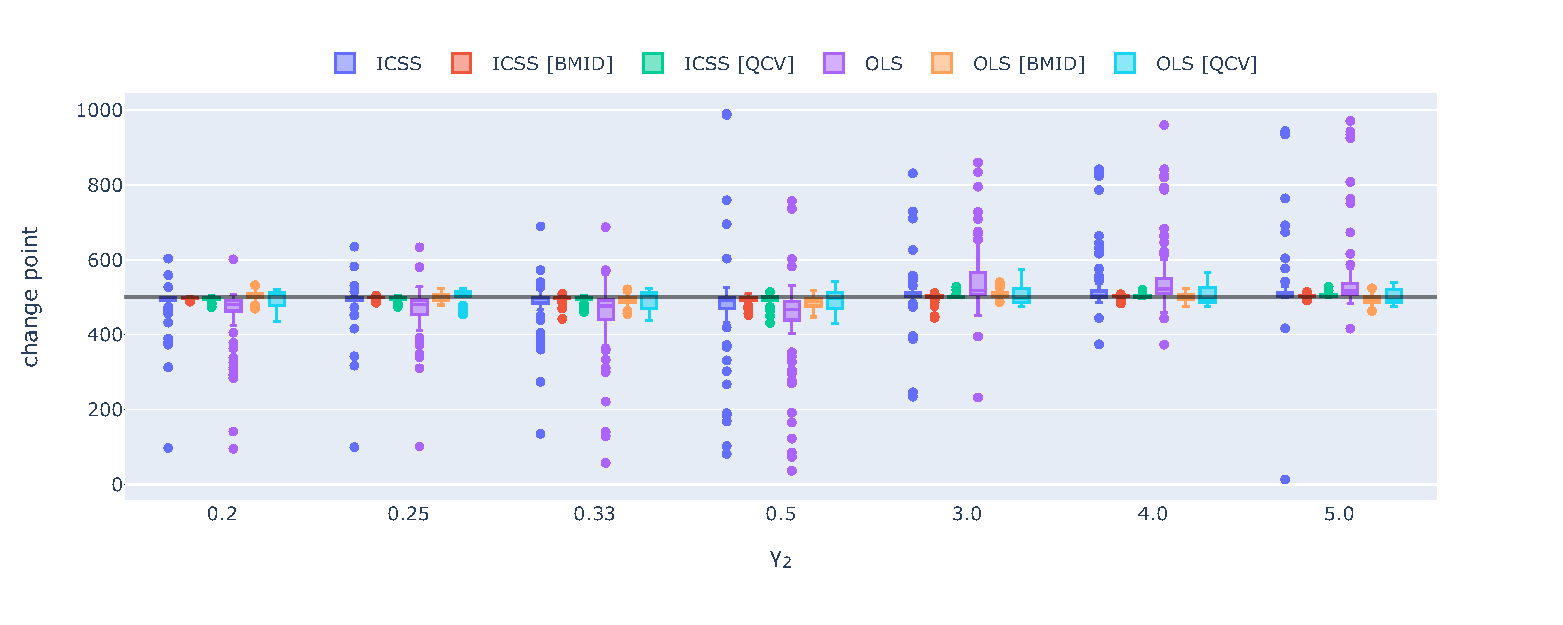}
    \caption{Detected change points obtained using selected procedures for sample described by model \eqref{eq:formula_stable} for $N=1000$. Distribution of the data: symmetric $\alpha-$stable distribution with \uline{$\alpha=1.9$}. True change point is marked via grey line.}
    \label{fig:sim_as_alpha19}
\end{figure}

\begin{figure}[h!]
    \centering
    \includegraphics[width=0.95\textwidth]{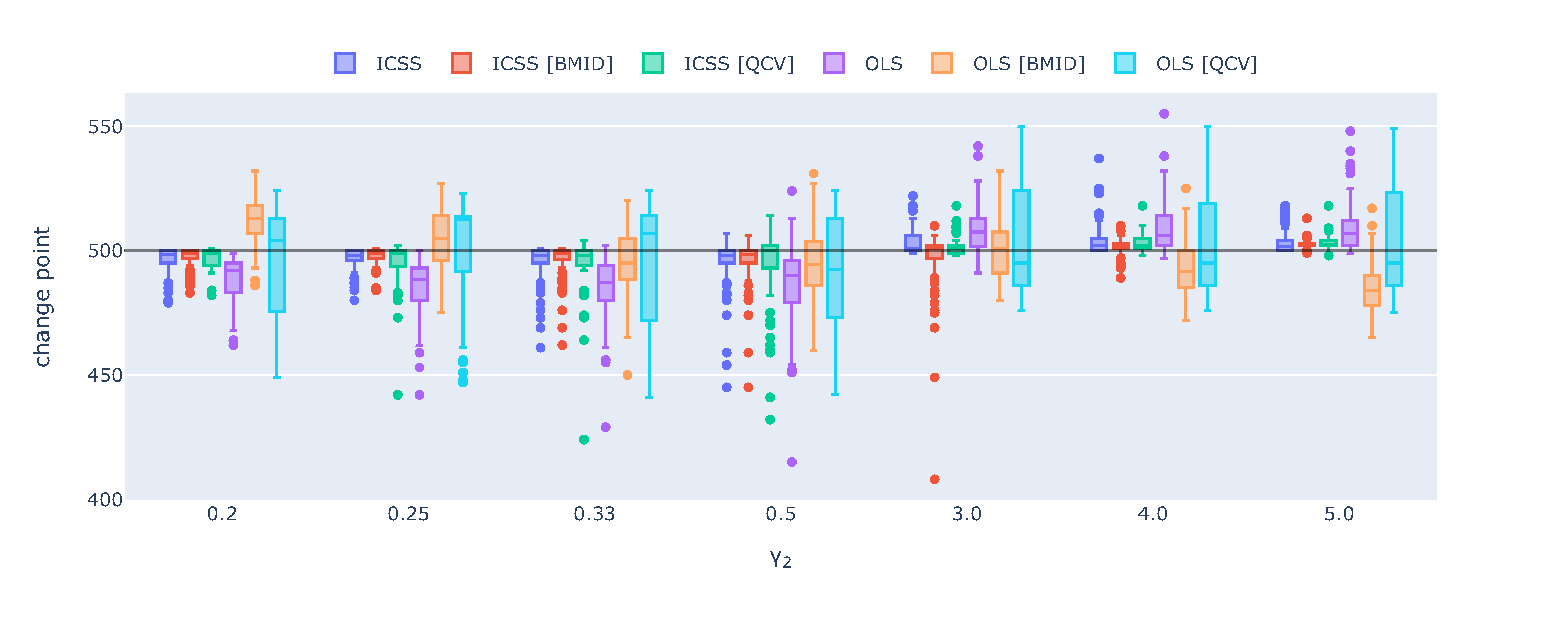}
    \caption{Detected change points obtained using selected procedures for sample described by model \eqref{eq:formula_stable} for $N=1000$. Distribution of the data: symmetric $\alpha-$stable distribution with \uline{$\alpha=2$}. True change point is marked via grey line. }
    \label{fig:sim_as_alpha2}
\end{figure}

The effectiveness of the proposed approaches for change point detection is tested in Monte Carlo trials. For each value of $\alpha \in \{1.1, 1.2, 1.3, ..., 1.9, 2.0\}$ and for various values of $\gamma_2$ we simulate the data given by the model \eqref{eq:formula_stable} in 100 trials. {However, the simulation results we present here only for selected values of $\alpha$ (i.e., $\alpha=1.1$, $\alpha=1.9$, and $\alpha=2$).} The change point is detected using baseline methods (ICSS and OLS procedures) and their robust counterparts. The illustrative results (i.e., detected change points) for $N=1000$ are presented in the boxplots (Fig. \ref{fig:sim_as_alpha11} -- Fig. \ref{fig:sim_as_alpha2}).

The results for sample size of $N=250$ are demonstrated in the Appendix (Fig. \ref{fig:sim_as_alpha11_n250} -- Fig. \ref{fig:sim_as_alpha2_n250}). It can be seen that robust methods detect change points in a more accurate way, especially for small values of $\alpha$. For $\alpha=1.1$ (Fig. \ref{fig:sim_as_alpha11}), we can observe that the medians of the detected change points for the robust methods are close to the gray line (i.e. the true change point). Conversely, for baseline approaches, we obtain significantly worse results: medians of the detected change points are far from the theoretical level. Moreover, for baseline methods, the boxplots have a higher inter-quartile range. For $\alpha = 1.9$ (Fig. \ref{fig:sim_as_alpha19}), the medians of the detected change points are close to the theoretical value for each of the considered methods. However, for robust procedures, a small number of outliers can be observed (unlike the baseline methods for which the number of outliers is relatively large). It is worth noting that for a Gaussian distribution ($\alpha=2$), the results for robust methods are comparable to the results received for baseline approaches.

In order to demonstrate that the discussed robust algorithms outperform the classical methods, in Tab. \ref{tab_new1} we present the mean absolute error (MAE) \citep{fomby2008scoring} calculated for the estimated change points for the data from symmetric $\alpha-$stable distribution for three $\alpha$ values discussed above and $N=1000$. The MAE is calculated based on 100 simulated samples.  The obtained results clearly show that the proposed robust methods return significantly smaller MAE values than the baseline algorithms, especially for the non-Gaussian case. {We can also notice that the smaller the $\alpha$ parameter, the greater the differences between the baseline approaches and the proposed robust methods. In extreme cases, the MAE for robust methods can be almost 19 times lower than for the baseline approaches.}

\subsection{Mixture Gaussian model}
\label{sec:mix_gauss_distr}
As a second example we consider the random sample $X_1,X_2,...,X_N$ satisfying the following model:
\begin{align}
    X_n \stackrel{d}{=} G_n + U_n \cdot K_n, \quad n=1,2,...,N,
    \label{eq:mix_gauss_def}
    \end{align}
\vspace{-0.1cm}
\vspace{-0.2cm}
where 
\begin{enumerate}
    \item $G_n$ is a random variable defined as follows: 
    \begin{align}
    G_n \stackrel{d}{=} \begin{cases}
        G^{(1)} \sim N(0, \omega_1^2=1) & \text{for } n < n^*, \\
        G^{(2)} \sim N(0, \omega_2^2) & \text{for } n \geq n^*. 
        \label{eq:G_n}
    \end{cases}
\end{align}
In the above equation,  $n^*$ is a change point and $\omega_2 \neq 1$.
    \item $U_n$ is an uniformly distributed random variable such that for each $n=1,2,...,N$, $U_n \stackrel{d}{=} U(0, \nu)$, $\nu=constant$, 
    \item  $K_n$ is bimodal random variable such that for each $n=1,2,...,N$, $P(K_n=1)=P(K_n=-1)=p, \quad P(K_n=0)=1-p$. Here $p \in (0, 1)$ describes probability of positive or negative outlier. 
\end{enumerate}

\begin{figure}[h!]
    \centering
    \includegraphics[width=0.9\textwidth]{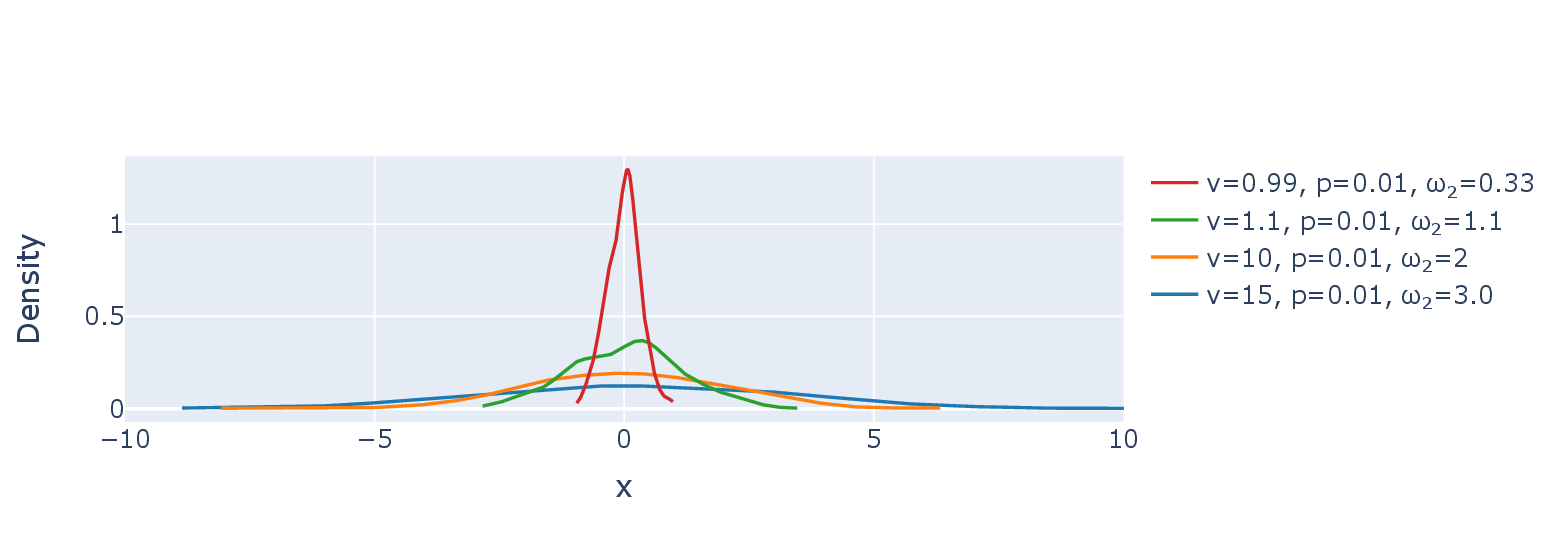}
    \caption{The p.d.f.s for the mixture of Gaussian model (second segment) for different values of the parameters.
    }
    \label{fig:ex_mix_gauss}
\end{figure}

We assume that the random variables $G_n,K_n$ and  $U_n$ are independent. Let us note that $K_n$ and $U_n$ refer to the probability and amplitude of an individual disorder, respectively. The p.d.f. of the exemplary distributions corresponding to the model \eqref{eq:mix_gauss_def} for the second segment are presented in Fig. \ref{fig:ex_mix_gauss}. The corresponding trajectories (i.e., simulated data from distributions presented in Fig. \ref{fig:ex_mix_gauss}) are shown in the Appendix (see Fig. \ref{fig:trajectory_mix_1}--\ref{fig:trajectory_mix_4}).

\begin{figure}[h!]
    \centering
    \includegraphics[width=0.95\textwidth]{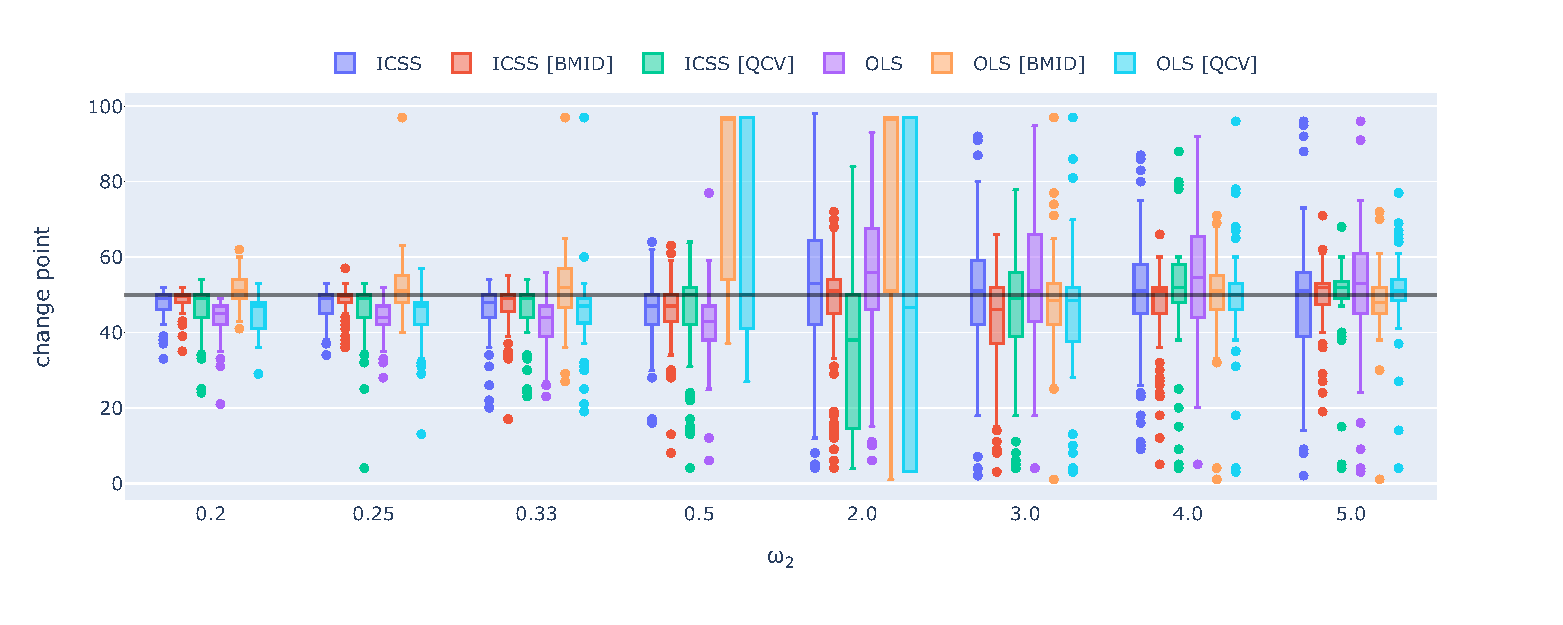}
    \caption{Detected change points obtained using selected procedures for sample described by \eqref{eq:mix_gauss_def} for $N=1000$. Distribution of the data: the mixtured Gaussian distribution for particular parameters: $p=0.05, \quad \nu=5\cdot \omega_2$.}
    \label{fig:mix_gaussA}
\end{figure}

\begin{figure}[h!]
    \centering
    \includegraphics[width=0.95\textwidth]{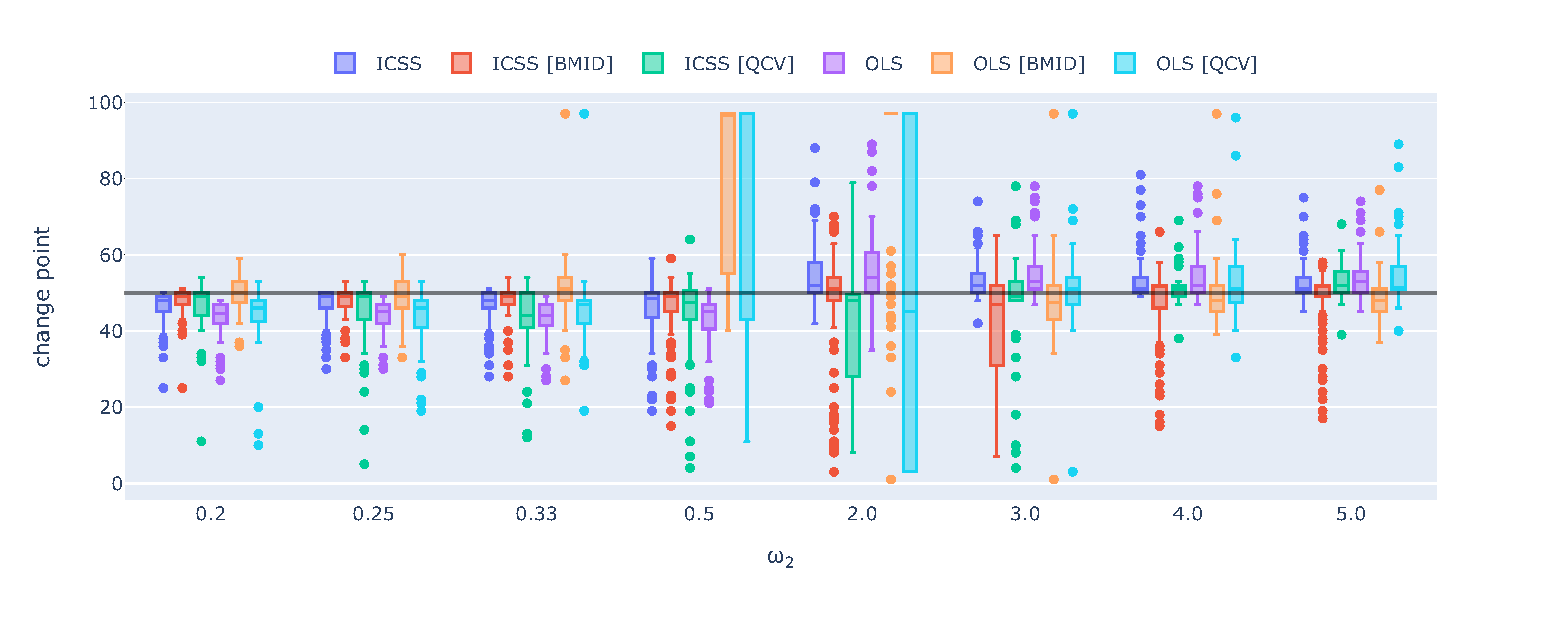}
    \caption{Detected change points obtained using selected procedures for sample described by \eqref{eq:mix_gauss_def} for $N=1000$. Distribution of the data: the mixtured Gaussian distribution for particular parameters: $p=0.05, \quad \nu=1.5\cdot \omega_2$.}
    \label{fig:mix_gaussB}
\end{figure}

For various hyperparameters space (i.e., for different $p, \omega_2, \nu$ values), we conduct 100 Monte Carlo simulations. The procedure is the same as in Section \ref{sec:alpha_stable_distr}. Figs. \ref{fig:mix_gaussA} and \ref{fig:mix_gaussB} present the results of the illustrative simulations. As we can see, for $\nu=5$ and  $\omega_2 \geq 2$: the robust approaches outperform the baseline algorithms (Fig. \ref{fig:mix_gaussA}). A median of each boxplot is close to theoretical value of change point, but the boxplots corresponding to baseline methods have wider range than boxplots of detected change points using robust methods. If $\nu$ is not much different than $\omega_2$ (i.e., $\nu/\omega_2 < 2$), then we can observe that results depends on the selected robust estimator of scale and selected method (ICSS or OLS).

When $\nu=1.5\cdot \omega_2$ and $\omega_2\neq 2$ then the median in the boxplots for OLS-based approach is closer to the true change point for robust procedures (Fig. \ref{fig:mix_gaussB}). For ICSS, the results of all methods are similar. This is because the outliers from the distribution $U(0, \nu=1.5\cdot \omega_2)$ are not very large relative to the cleaned data (without the presence of outliers). {Similar as in  Section \ref{sec:alpha_stable_distr}, we demonstrate the advantage of robust methods over baseline ones using MAE. Tab. \ref{tab_new2} shows MAE calculated for the detected change points for the data from the mixture Gaussian model. Here we consider the hyperparameters discussed above, that is, $\omega_2 \in \{ 0.2, 0.25, 0.33, 0.5, 2, 3, 4, 5\}$ and $\nu  \in \{ 1.5 \cdot \omega_2, 5 \cdot \omega_2 \}$. The sample size for each of the 100 Monte Carlo trials is $N=1000$. If noise-related peaks have large values ($\nu = 5\cdot \omega_2$) then robust methods outperform the classical ones. However, there are differences in performance with respect to the selected robust scale estimator. For the lower peaks ($\nu = 1.5 \cdot \omega_2$), the behavior of the baseline and robust methods is similar. This is consistent with intuition (i.e., robust methods should be more effective if the data is heavy-tailed distributed).}

\subsection{Time complexity of the proposed algorithms}
\begin{figure}[h!]
    \centering
    \includegraphics[width=0.7\textwidth]{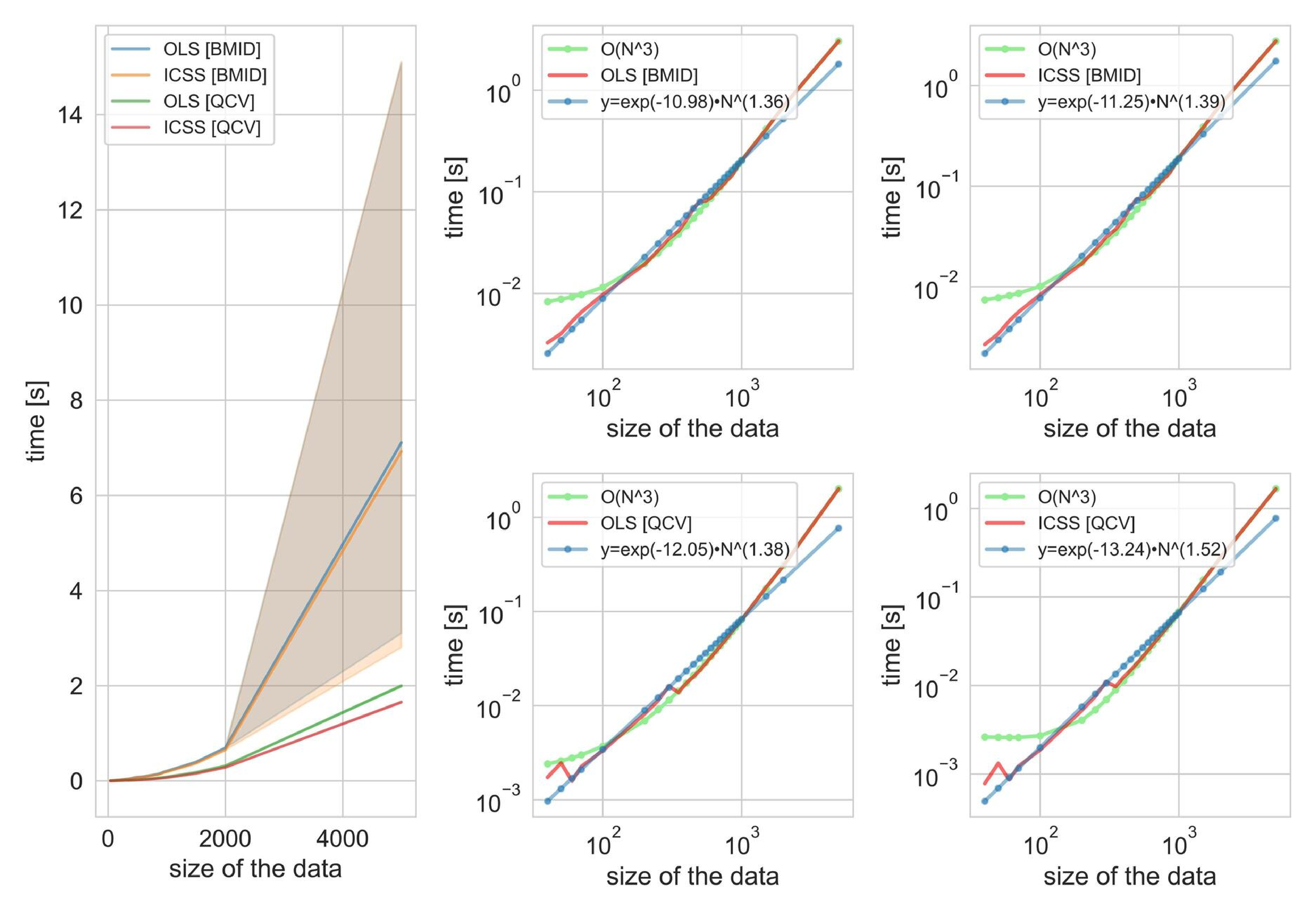}
    \caption{Analysis of computational time of the proposed procedures for data from model \eqref{eq:mix_gauss_def}.}
    \label{fig:complexity}
\end{figure}

The last step of the simulation study is to examine the time complexity of the proposed robust approaches. Fig. \ref{fig:complexity} presents the time of change point detection for robust versions of ICSS and the OLS-based methodd. The time is calculated for simulated data corresponding to the model \eqref{eq:mix_gauss_def} 
where 
$\nu = 15 , \omega_2 = 5$, \\ $p = 0.05, N \in \{40,   50,   60,   70,  100,  200,  250,  300,  350,  400,  450,
        500,  550,  600,  650,  700,  750,  800,  850,  900, 950, \\ $ 1000,
       $1500, 2000, 5000  \}$ and $n^*$ is located in the middle of the data.

The plot on the left-hand side shows the comparison of computational time for considered robust approaches. More precisely, for each $N$ we calculate the median of computational time obtained in 100 Monte Carlo trials. Furthermore, we present the confidence intervals for the computational times at the significance level $0.95$. The confidence intervals for \textit{OLS [QCV]} and \textit{ICSS [QCV]} are narrow, therefore they are not easily visible. As can be seen, the algorithms \textit{ICSS [QCV]} and \textit{OLS [QCV]} are much faster than their equivalents that use \eqref{eq:BMID} as a scale estimator (that is, \textit{ICSS [BMID]} and \textit{OLS [BMID]}). It means that the type of robust scale estimator may influence the computation time (depending on the properties of the data distribution). In the four graphs on the right-hand side of Fig. \ref{fig:complexity} we present the median computational time obtained in 100 Monte Carlo simulations for each value of $N$ separately for each of the considered algorithms. The \textit{ICSS [BMID]}, \textit{OLS [BMID]} and \textit{OLS [QCV]} behave like $f(N) \approx c \cdot N^{1.37}$. %The constant $c$ has a higher value than the coefficients of third-degree polynomials (for \textit{OLC [QCV]} and \textit{ICSS [QCV]}), thus $\hat{\sigma}^2_{r1}$-based methods seems to be faster for much larger values of $N$ than those considered in this paper. %It can be seen that the time complexity of the particular procedure depends on the selected robust estimators of scale.

The analogous figures corresponding to model \eqref{eq:formula_stable} are presented in the Appendix (Fig. \ref{fig:complexity_alpha_stable}). For simulations, we take $\alpha = 1.5, \gamma_2=5$ and $N$ from the same set as for the previously presented experiment in the case of the mixture Gaussian model. For symmetric $\alpha$-stable distributed data, the time complexity of ICSS and OLS is similar regardless of the robust estimator of scale. In this case, the slight differences resulted from the characteristics of the data distribution. It is worth noting that, for other distributions, the choice of a robust scale estimator may be crucial in the context of optimizing the time of finding the change point. A good example of a distribution for which we observed this relationship is the Gaussian mixture model mentioned earlier.

\section{Real data analysis}
\label{sec:real_data}
In this section, we demonstrate the effectiveness of the proposed methods of change point detection in the real-world applications. The analyzed data are from three sectors: financial, industrial, and medical. 

\subsection{The logarithmic exchange rate}

The first data set comes from the financial sector. Here, we analyze the returns on dollars and the Polish zloty exchange rate (USDPLN). In our analysis, we take logarithmic returns.  The same data was analyzed in \citep{bielak2021market}, where the authors indicated the existence of a connection between the USDPLN exchange rate and the financial risk of the Polish mining company. In \citep{bielak2021market}, Hidden Markov Model (HMM) was used in the process of classifying states occurring in the trajectory of logarithmic returns. This method was used to divide the data into segments. The obtained change point agrees with the actual interpretation. In 2008-2009, there was a crisis that affected the dynamics of many financial instruments. \textit{''The end of regime 1 is specified as the second half of 2012 when the situation on the market has started to stabilize and the classification results indicate the second regime for both assets''}, as suggested by the authors of \citep{bielak2021market}.

\begin{figure}[h!]
    \centering
    \includegraphics[width=0.95\textwidth]{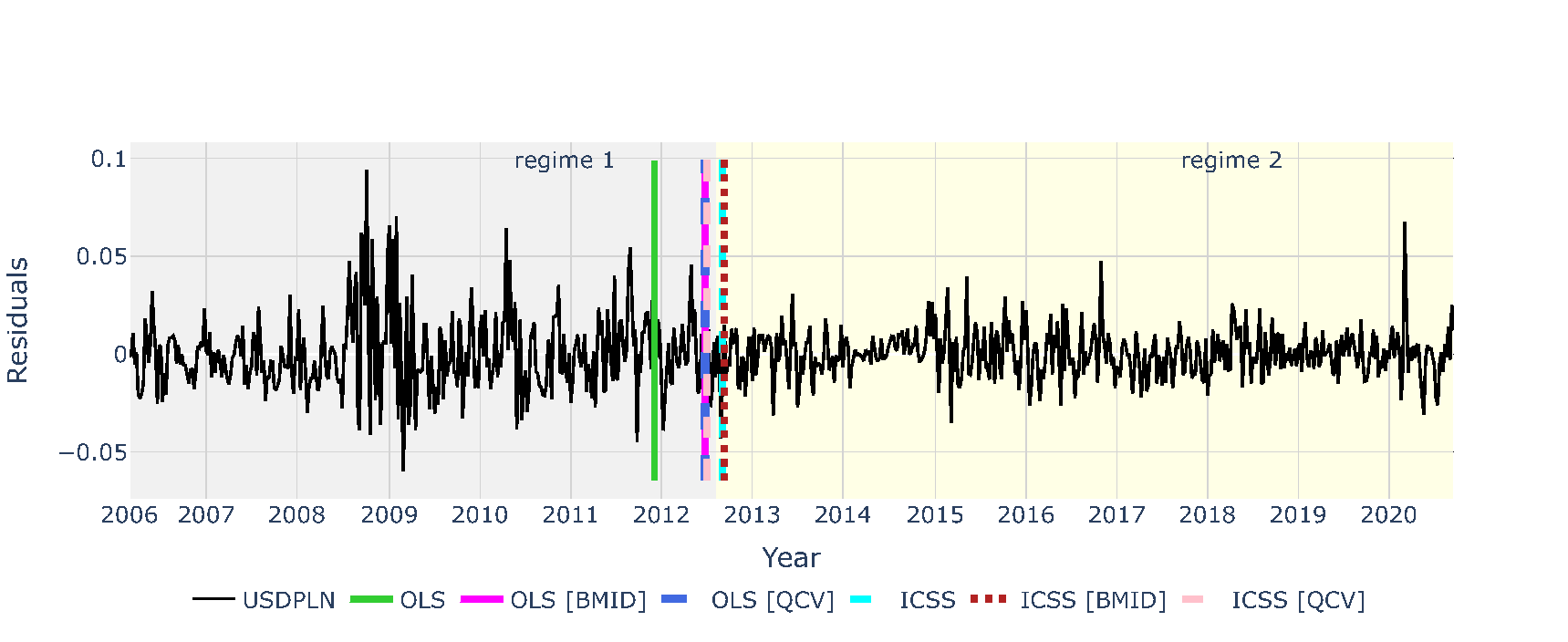}
    \caption{The segmentation results for the financial data. The residuals corresponding to the VAR(1) model applied to the logarithmic returns of the weekly USDPLN exchange rate. The methodology of modeling is described in \citep{bielak2021market}.}
    \label{fig:real_finance}
\end{figure}

Financial data are often characterized by long-tail characteristics (heavy-tailed behavior). In \citep{bielak2021market} it was shown that both segments of the data set come from the $\alpha$-stable-distributed model with parameter $\alpha$ close to $2$. In our analysis, we applied the discussed algorithm to confirm the results obtained using HMM. The segmentation results are visible in Fig. \ref{fig:real_finance}. Most of the discussed methods work correctly: the change points (represented by vertical lines) are close to the point where the data regime changes (background color changes from gray to yellow). As one can see, the use of robust scale estimators improved the results for OLS algorithm, while the ICSS technique performed well in all variants (baseline and robust). The fact that $\alpha$ was close to $2$ was critical here.

To present the differences between the algorithms quantitatively, we calculate the normalized error that is defined as follows: 
\begin{align}
    error = \frac{|n^* - {\hat{n}^*}|}{N},
    \label{eq:real_data_error}
\end{align}
where $\hat{n}^*$ is a detected change point,  $n^*$ is a true change point, and $N$ is a sample size. The values of normalized errors are presented in the Tab. \ref{tab:finance_errors}. The error of the OLS method is an order of magnitude higher than the errors of its robust versions. {Importantly, according to the results presented in \citep{bielak2021market}, the data from both segments come from the $\alpha$-stable distribution with the $\alpha$ parameter close to 2. The small error for the baseline ICSS method is due to the fact that the data in both segments resemble close to Gaussian distribution.}

\begin{table}[h!]
\centering
\caption{The normalized error calculated using \eqref{eq:real_data_error} for financial data. Sample size $N=760$. }
 \label{tab:finance_errors}
\begin{tabular}{cccccc}
\hline
\multicolumn{1}{r}{ICSS} & \multicolumn{1}{r}{ICSS {[}BMID{]}} & \multicolumn{1}{r}{ICSS {[}QCV{]}} & \multicolumn{1}{r}{OLS} & \multicolumn{1}{r}{OLS {[}BMID{]}} & \multicolumn{1}{r}{OLS {[}QCV{]}} \\ \hline
0.00526                  & 0.00658                             & 0.00658                            & 0.04605                 & 0.00789                            & 0.00789                           \\ \hline
\end{tabular}
\end{table}

\subsection{The vibrations measurements}
The second real-world problem considered in our research is the segmentation of vibration measurements. We use three sensors in the data aggregation process: a linear accelerometer, a gravity force meter, and a gyroscope. The measuring device is a smartphone that uses the application \textit{Physics Toolbox Sensor Suite} that is available for Android. The basic characteristics of these sensors are given in Tab. \ref{Tab:sensors}. 

\begin{table}[tbh!]
\centering
\caption{\label{Tab:sensors}Basic paremeters of the sensors.}
\begin{tabular}{lccc}
\hline
\text{Sensor}                                                              & G-Force Meter & Gyroscope        & Linear Accelerometer \\ \hline
\text{Range}                                                               & $\pm 4g$                & $\pm 17.45 rad/s$ & -            \\ 
Maximum sampling frequency & 100Hz                   & 100Hz           & 100Hz                \\ \hline
\end{tabular}
\end{table}

Nowadays, there is an increasing demand for algorithms that support automatic processing of signals from these sensors. This is directly related to the increase in the number of devices that use them to self-identification of their state. An example of a device is a smartwatch that automatically sets the movement mode based on the level and frequency of vibrations. Another example is a laptop that turns off in case of unexpectedly high vibrations, which can be associated with a possible fall of the device. It reduces the chance of damaging the device's components during a fall is reduced.

\begin{figure}[h!]
    \centering
    \includegraphics[width=0.95\textwidth]{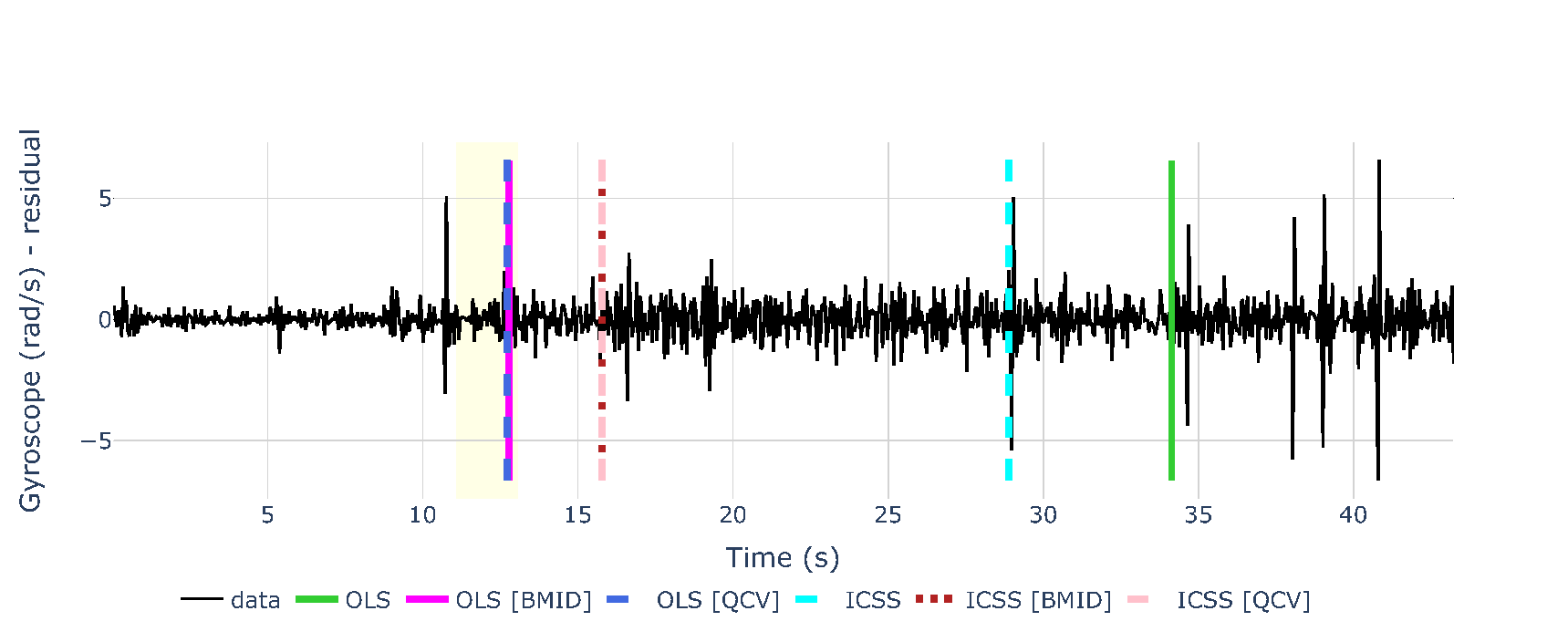}
    \caption{The segmentation result for the data from accelerometer.}
    \label{fig:real_gyro}
\end{figure}

The data acquisition process includes recording the oscillation induced by a human or vehicle (car/motorbike). More details on the data can be found in \citep{ww2022identification}. Intuitively, the gravitational force or angular velocity might exhibit a relatively constant variance under specific unchanging conditions. However, certain environmental conditions or the dynamics of the facility may affect the vibration level. In \citep{ww2022identification}  two experiments were described in which vibrations are aggregated during a change in a specific condition.

In this research, we use a sample recorded for motorbikes that are moving from one place to another. The change in the surface (from an asphalt to a stone route) is the factor that causes the dynamics of variance. The results of the segmentation for these data are presented in Fig. \ref{fig:real_gyro}. Data are aggregated and differentiated. Differentiation is performed to remove dependencies present in the signal. As one can see, the robust versions of the OLS technique segmented the data in the most accurate way. The detected change points are in the yellow range, that is, the range where the true change point is located. The range was determined based on the course of the experiment by an independent observer. 

Importantly, robust versions of ICSS also outperform the baseline methods. The detected change points are only a few seconds after the true change point, whereas baseline methods estimate changes several dozen seconds away from the true change point. The error values are given in the Tab. \ref{tab:acc_errors}. In this case, $y_2$ is taken as the median of the points that bound the yellow range. It can be seen that robust methods are much more effective than baseline algorithms.  

\begin{table}[h!]
\centering
\caption{The normalized error calculated using \eqref{eq:real_data_error} for vibration data. Sample size $N=860$. }
 \label{tab:acc_errors}
\begin{tabular}{cccccc}
\hline
 ICSS     & ICSS {[}BMID{]} & ICSS {[}QCV{]} & OLS      & OLS {[}BMID{]} & OLS {[}QCV{]} \\ \hline
 0.389 & 0.085        & 0.085       & 0.510 & 0.016       & 0.015      \\ \hline
\end{tabular}
\end{table}

\subsection{Medical data - acoustic body sounds}

Living organisms are complex systems with multiple entangled processes, originating at the chemical-physical level of cells, determining organization through tissues and organs, which in turns triggers the emergence of a wealth of behaviors up to the whole-body level, difficult for continuous monitoring and control. Let us take only the example of a brain encephalography (ECG) measurement procedure which provides outputs in the form of complex noisy-like patterns with spiky events. Thousands of works have been published in the last decades that try to decipher useful information from these complex data. The goal is not only to understand the physiology of human beings, but also to use these time series data in human-machine interaction \citep{Song_2020}. The same is with the other channels recorded during polysomnography, like electromyography (EMG), oculogram, leg actigraphy, etc. \citep{Liang_2019}.

In the paper, we provide an example with unexplored patterns of acoustic body sounds measured with a microphone at the lungs level in COVID-19 patients. In \citep{Ranta_2010} the authors have shown that differentiation between patterns from body sounds measured in digestive tract can provide new insight into medical diagnostics of the gastrointestinal motility disorders. Sheu et al. in \citep{Sheu_2015} improved the procedure with the use of higher-order statistics modeling  and in \citep{Zhao_2020} it was  demonstrated how bowel sounds - recorded with a portable device - can be segmented by convolutional neural networks. Additionally, Bondareva et al. in \citep{Bondareva_2022} have shown that it is possible to disentangle even stress level from body sound data measured at the abdominal level. In general, research and engineering efforts are reported focused on the design of a digital stethoscope, which can allow easy and non-invasive monitoring of chronic diseases in home conditions \citep{Malik_2017, Kwiatkowski_2023}. The COVID-19 pandemy highlighted the next advantage of simple lung tests based on spontaneous respiration - risk minimization of virus spreading \citep{Baretto_2020, BEYDON_2020}. On the other hand, time series recording of the acoustic phenomena measured during spontaneous breathing at the level of the chest can be biased with natural and unintentional components of various distributions, e.g. associated with cough, wheezing, snoring, crackles from lungs and generated in surroundings, talk, sensor displacement on the body surface, etc. The issue is to design a fully automatized stethoscope, which means that the signal processing procedure should be able to differentiate between useful and noisy components, providing a reliable diagnosis. Finally, data labeling should be performed for measured data and the procedure need to identify the moment in a sequence of measured samples when the working regime changes between two various states of the physiological system operation.

For the purpose of testing the novel data segmentation algorithm, acoustic sounds were measured in hospitalized patients with COVID-19 with the device used also in \citep{Bondareva_2022}. Measurement sessions were approved by the Bioethical Commission of the Wroclaw Medical University - no. of the decision 642/2021. Written informed consent was obtained from all participants. The patients were in a sitting position. The microphone was mounted in a selected place on the body at chest level in two positions at the front and four positions at the back. The sampling rate applied for the recorded data was 44.1 kHz. Before registration patients were auscultated by a medical doctor using an analog stethoscope. The recording was simultaneously monitored in real time via headphones. The measurement covered the entire respiratory cycle, inhalation and exhalation, and lasted two minutes during both spontaneous and deep breathing. The results of auscultation were used for triage and then recorded data were labelled in details using digital recordings.

\begin{figure}[h!]
    \centering
    \includegraphics[width=0.95\textwidth]{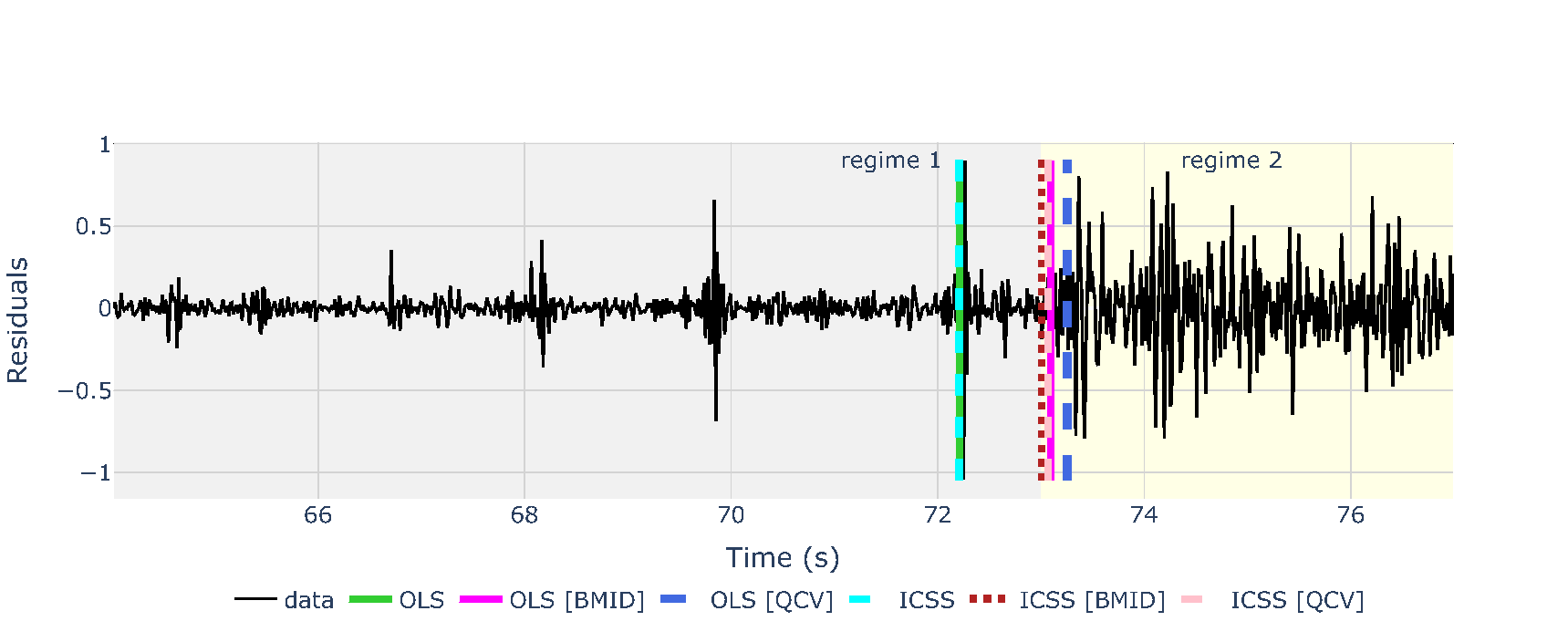}
    \caption{The segmentation result for the medical data. Change point is connected with the environmental noise.}
    \label{fig:real_medical}
\end{figure}

Fig. \ref{fig:real_medical} shows the result of segmentation for medical data for which the scale change point is related to the occurrence of disturbances in the environment. Before segmentation, the data was aggregated (to reduce resolution) and differentiated. The baseline for the identification of a true point of change was determined listening to the recordings by the medical doctor. It can be seen that robust methods predict the change point more accurately. They are insensitive to the occurence related to the respiratory cycle.

Tab. \ref{tab:medical_errors} shows that the errors of the baseline methods (OLS and ICSS) are an order of magnitude higher than the errors of their robust versions.

\begin{table}[h!]
\centering
\caption{The normalized error calculated using \eqref{eq:real_data_error} for medical data. Sample size $N=990$. }
 \label{tab:medical_errors}
\begin{tabular}{cccccc}
\hline
ICSS  & ICSS {[}BMID{]} & ICSS {[}QCV{]} & OLS   & OLS {[}BMID{]} & OLS {[}QCV{]} \\ \hline
0.062 & 0.000           & 0.004          & 0.062 & 0.007          & 0.019         \\ \hline
\end{tabular}
\end{table}

Change point detection algorithms can also be used to segment data into more detailed patterns, e.g., the inspiratory and the expiratory cycles, and then differentiate between their details. With such segmentation,  data can be automatically grouped into inspiratory and expiratory cycles, and analyses dedicated to detecting intra- and inter-breath abnormalities correlated with the physiological context of lung disorders can be performed. Finally, long-term, reliable, and automatic monitoring is of particular importance in the management of chronic diseases (e.g. in lungs, digestive track, heart); thus designing robust procedures for raw data segmentation is fundamental and pre-step to more thorough analytical actions on acoustic, electric, or other nature signals. The reported study proves that we can reliably distinguish between acoustic regime changes in auscultated COVID-19 patients with the use of the robust OLS and the robust ICSS algorithms, regardless of the seed and origin of the impulsive and non-Gaussian distortions. The next step will be to provide an on-line mode to this procedure, operation on multiple change points and multivariate data, and also an extensive physical-mathematical modeling of the acoustic phemomena recorded during prolonged and digitized auscultation of patients suffering from lung diseases.  

\section{Discussion and Conclusions}\label{sec_Discussion}
\label{sec:discussion_conclusions}

We are on the eve of building autonomous systems where a measurement science and data-driven approach play fundamental roles. In fact, the concept of autonomy requires systematic monitoring and even prediction of the temporal state of a complex system and optimized management of available resources. All this means that tracking and understanding the evolution of a system (including its correct and anomalous operating regimes) should be algorithmized. 
Data segmentation is a data processing step that differentiates between changing regimes in an evolving system and is usually based on identifying the point of regime change.

Impulsive and non-Gaussian distortions in the measured data are significant challenges during data segmentation even for domain-experts, and the problem becomes even more difficult to algorithmize when the variance of the dataset goes to infinity.

The objective of this study is to propose a methodology suitable for the segmentation of non-Gaussian data with a time-changing scale. Our contribution consists in the introduction of offline algorithms which represent refined iterations of established techniques based on CSS statistics, namely ICSS and quantile methods. {The proposed approach is based on well-established methodologies, increasing their potential applicability in real-world data scenarios.

The robust versions of baseline methods for scale change point detection are validated for simulated data from two distributions, namely the  symmetric $\alpha$-stable distribution and the mixture Gaussian model. Both distributions belong to the class of heavy-tailed family, however they posse different properties. The goal was to create a  methodology dedicated for heavy-tailed distributed data with possible infinite variance (e.g., data from the $\alpha$-stable distribution with $\alpha < 2$), as well as data containing  extreme values (e.g., mixture Gaussian model). Using Monte Carlo trials, we compare the performance of baseline methods against their robust counterparts. The analysis of the distribution of the obtained change point estimations proves the superior efficacy of robust methods compared to heavy-tailed data originating from the $\alpha$-stable distribution and data arising from the Gaussian mixture model. {We have  shown that the MAE for robust methods can be up to 20 times smaller than the MAE for baseline techniques (Tab. \ref{tab_new1}) in most extreme case.  
Particularly, the largest difference in MAE is when the heavy-tailed properties are more visible in the data (e.g., for a $\alpha$-stable distribution with $\alpha=1.1$). If the data have a Gaussian (or close to Gaussian) distribution, then the effectiveness of the robust and baseline methods is comparable.}

The simulation studies are supported by analysis of real data with clear change points related to the analyzed phenomena. Specifically, in financial data, the change point is related to the 2008 economic crisis, manifesting itself as fluctuations in stock exchange dynamics. Within industrial datasets, the structural break point is precipitated by alterations in the surface terrain encountered during vehicular movement. Lastly, in the context of the medical dataset, the perturbation arises from disturbances in the patient's environment, reflecting alterations in observed physiological parameters. {The analysis show that for data exhibiting impulsive behavior the error in estimating the change point is significantly smaller for robust methods in comparison to  baseline algorithms.}

As a future study, we plan to expand the proposed robust methodology for multivariate data analysis. This approach is particularly important in financial data analysis, where the interplay between two or more assets requires a detailed understanding of scale changes, potentially pivotal for prognostic purposes. In such contexts, the assumption of Gaussian or finite-variance distributions may be inappropriate. Analogous concerns arise within medical data analysis, where the  examination of two or more physiological parameters offers more granular insight into anomalous conditions.

\section*{Acknowledgements}
\noindent
The work of A. Wyłomańska was supported by NCN grant under Sheng2 project No. UMO-2021/40/Q/ST8/00024 “NonGauMech - New methods of processing non-stationary signals (identification, segmentation, extraction, modeling) with non-Gaussian characteristics for the purpose of monitoring complex mechanical structures”.

\bibliography{mybibliography}

\appendix
\label{sec:appendix}
\section{The symmetric $\alpha$-stable distribution}

\vspace{-0.7cm}
\begin{figure}[h!]
    \centering
    \includegraphics[width=0.95\textwidth]{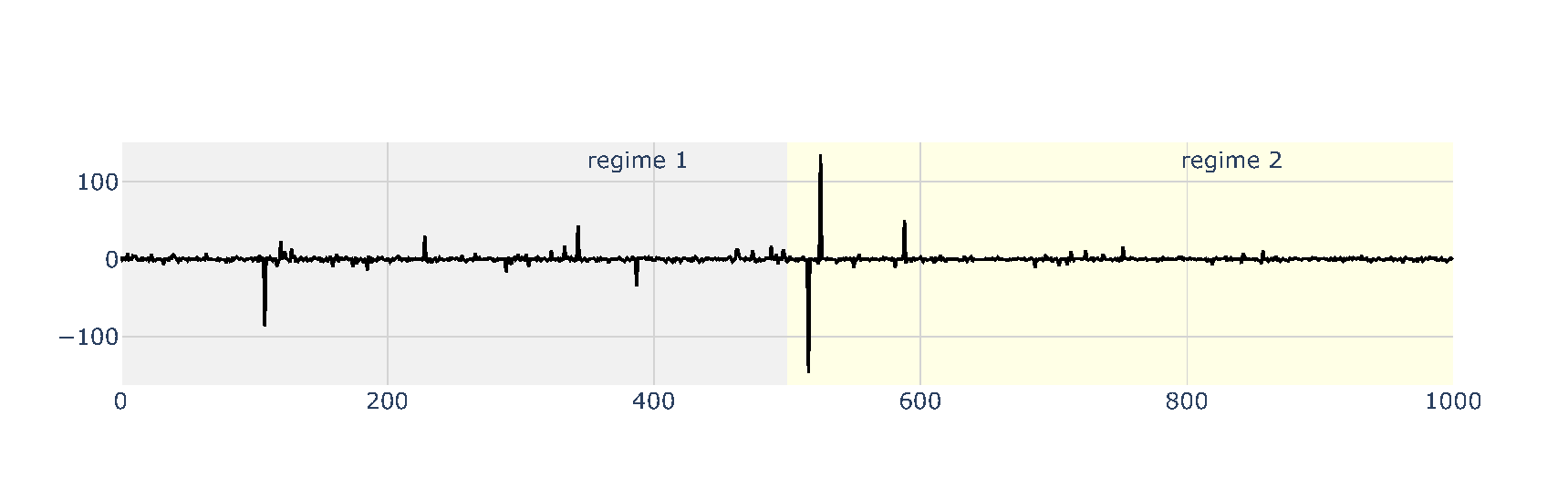}
    \caption{Example trajectory of data described by model \eqref{eq:formula_stable} for $\alpha=1.2, \gamma_1=1, \gamma_2=0.66.$}
    \label{fig:trajectory_alpha_12_gamma2_066}
\end{figure}

\vspace{-0.7cm}
\begin{figure}[h!]
    \centering
    \includegraphics[width=0.95\textwidth]{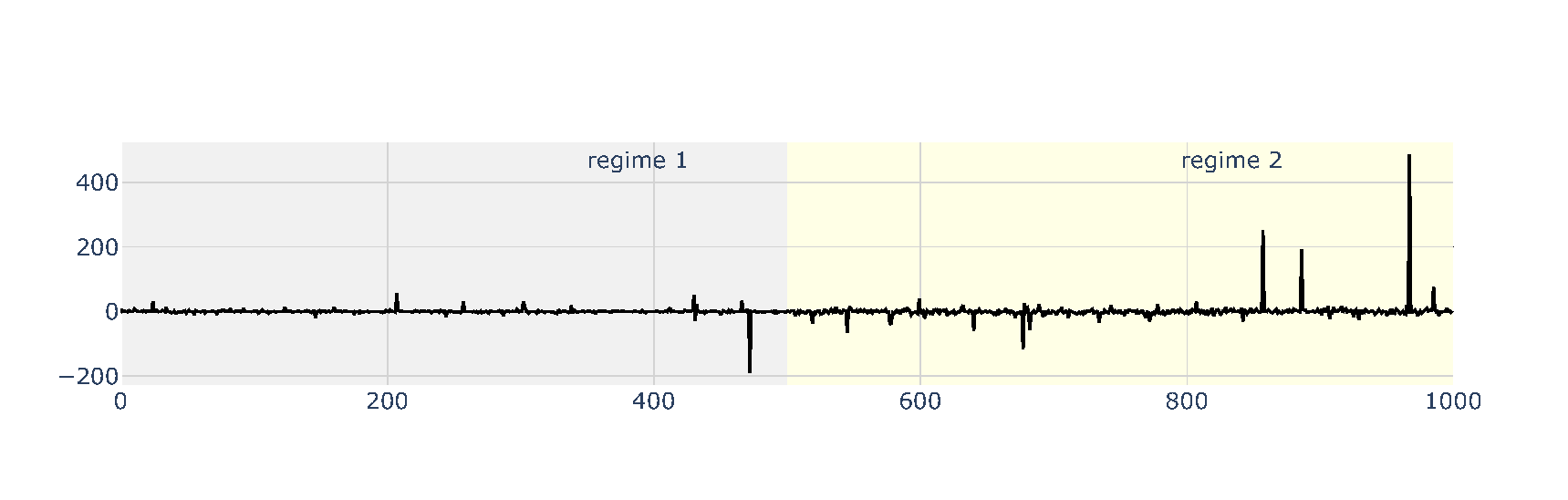}
    \caption{Example trajectory of data described by model \eqref{eq:formula_stable} for $\alpha=1.2, \gamma_1=1, \gamma_2=3.00.$}
    \label{fig:trajectory_alpha_12_gamma2_3}
\end{figure}

\begin{figure}[h!]
    \centering
    \includegraphics[width=0.95\textwidth]{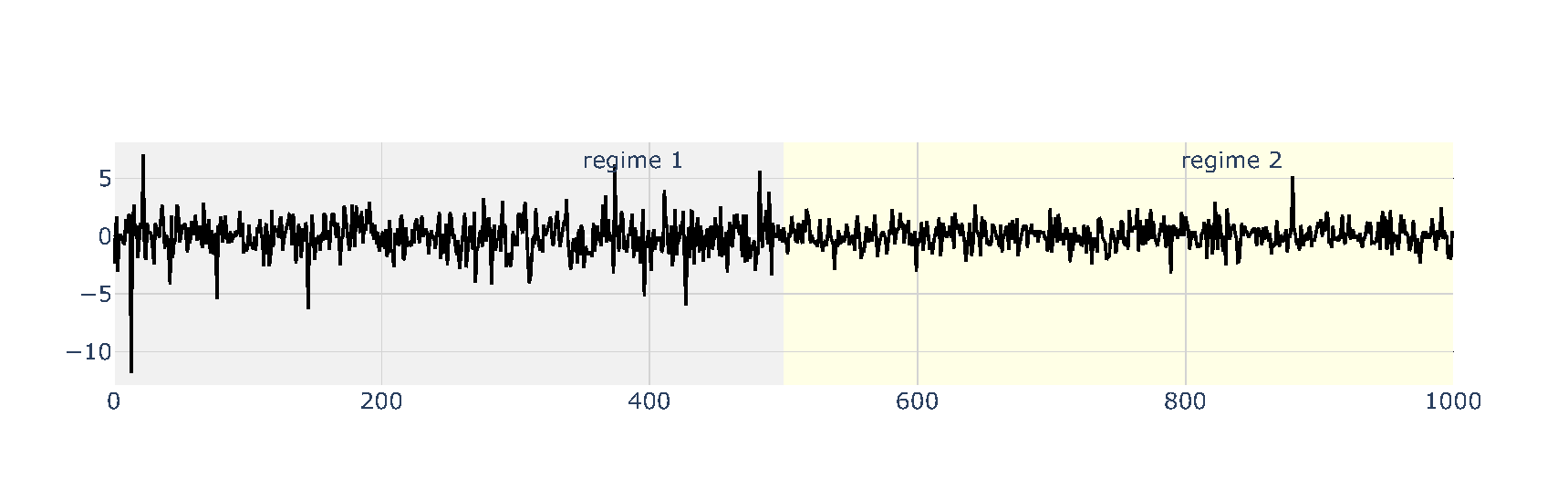}
    \caption{Example trajectory of data described by model \eqref{eq:formula_stable} for $\alpha=1.9, \gamma_1=1, \gamma_2=0.66.$}
    \label{fig:trajectory_alpha_19_gamma2_066}
\end{figure}

\begin{figure}[h!]
    \centering
    \includegraphics[width=0.95\textwidth]{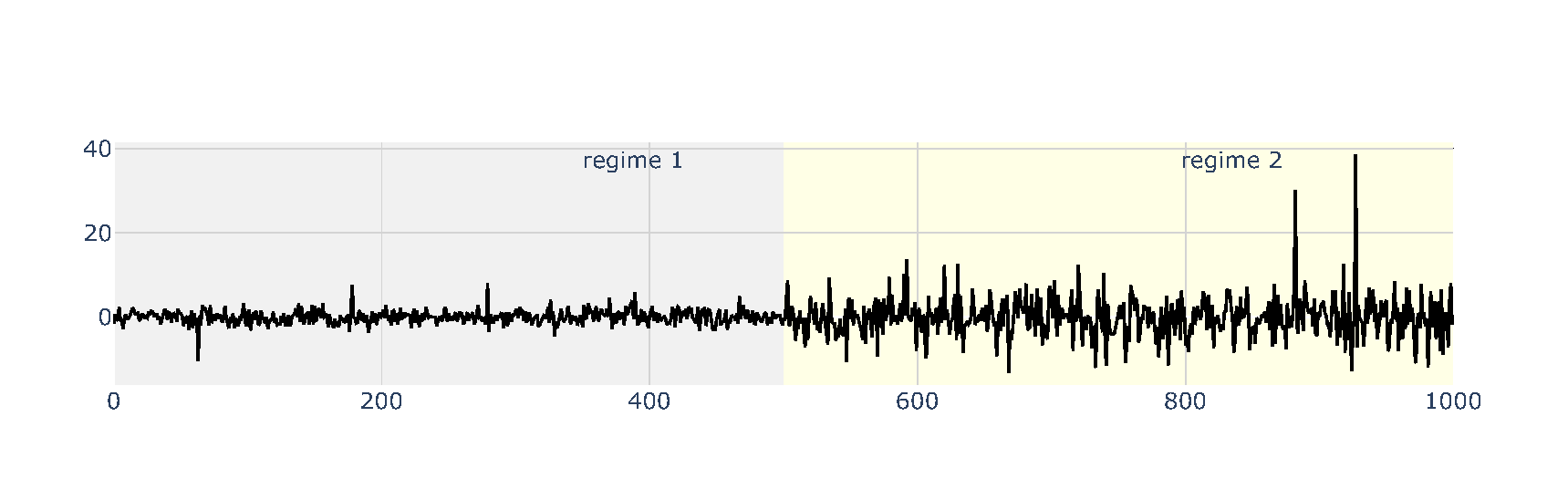}
    \caption{Example trajectory of data described by model \eqref{eq:formula_stable} for $\alpha=1.9, \gamma_1=1, \gamma_2=3.00.$}
    \label{fig:trajectory_alpha_19_gamma2_3}
\end{figure}

\begin{figure}[h!]
    \centering
    \includegraphics[width=0.9\textwidth]{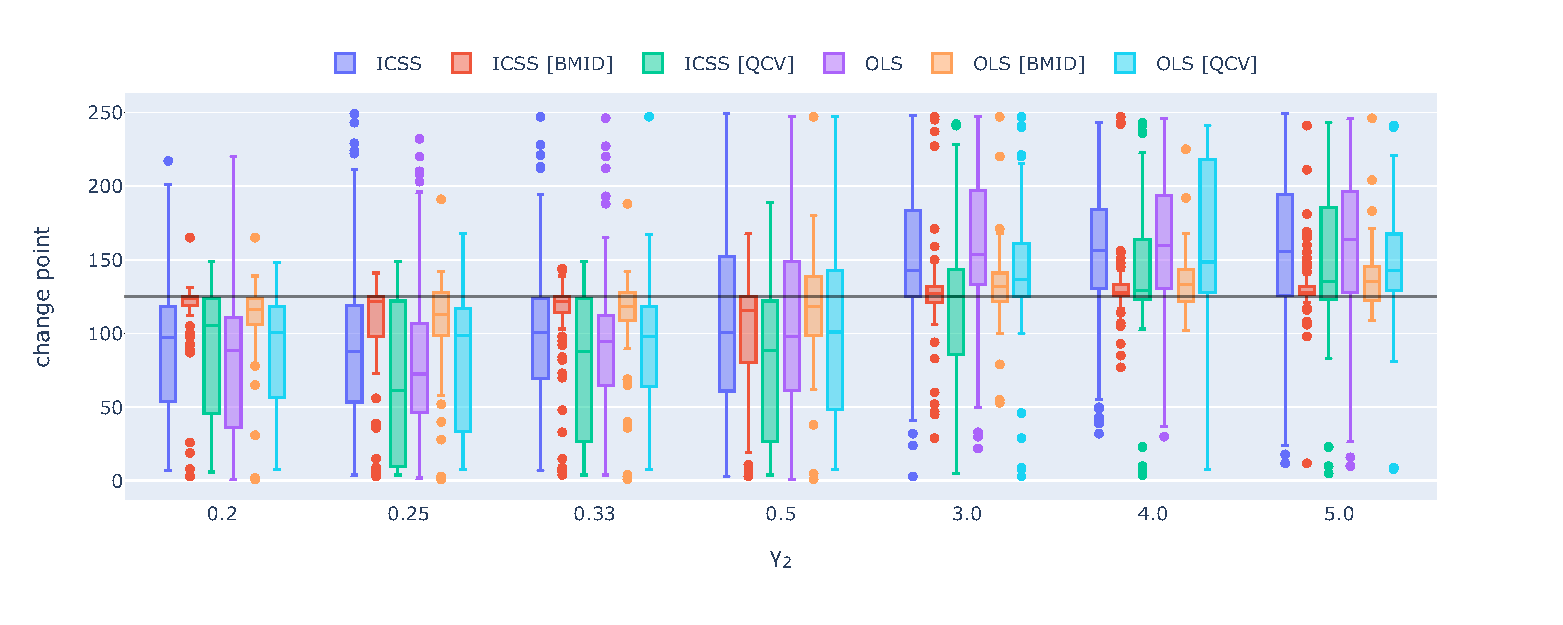}
    \caption{Detected change points obtained using selected procedures for sample described by model  \eqref{eq:formula_stable} with size $N=250$. Distribution of the data: symmetric $\alpha-$stable distribution with \uline{$\alpha=1.1$}. True change point is marked via grey line. }
    \label{fig:sim_as_alpha11_n250}
\end{figure}

\begin{figure}[h!]
    \centering
    \includegraphics[width=0.9\textwidth]{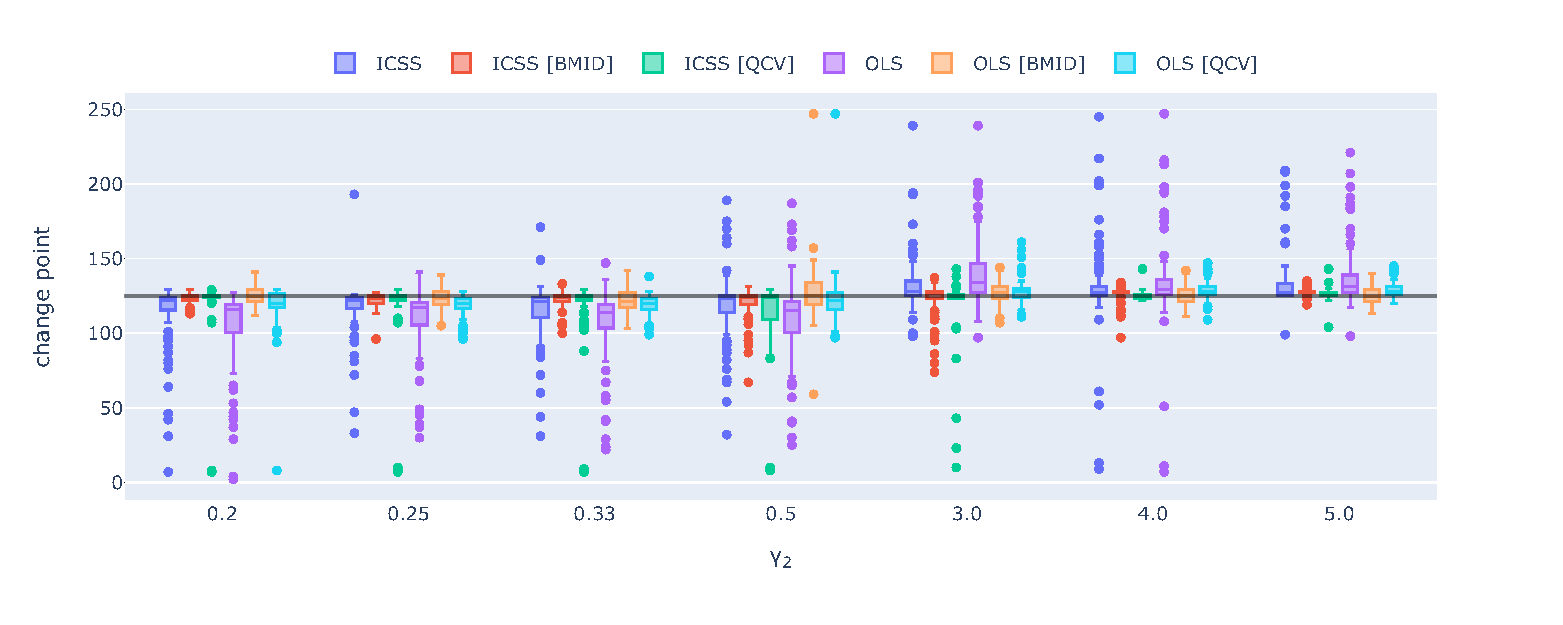}
    \caption{Detected change points obtained using selected procedures for sample described by model \eqref{eq:formula_stable} with size $N=250$. Distribution of the data: symmetric $\alpha-$stable distribution with \uline{$\alpha=1.9$}. True change point is marked via grey line. }
\end{figure}

\begin{figure}[h!]
    \centering
    \includegraphics[width=0.9\textwidth]{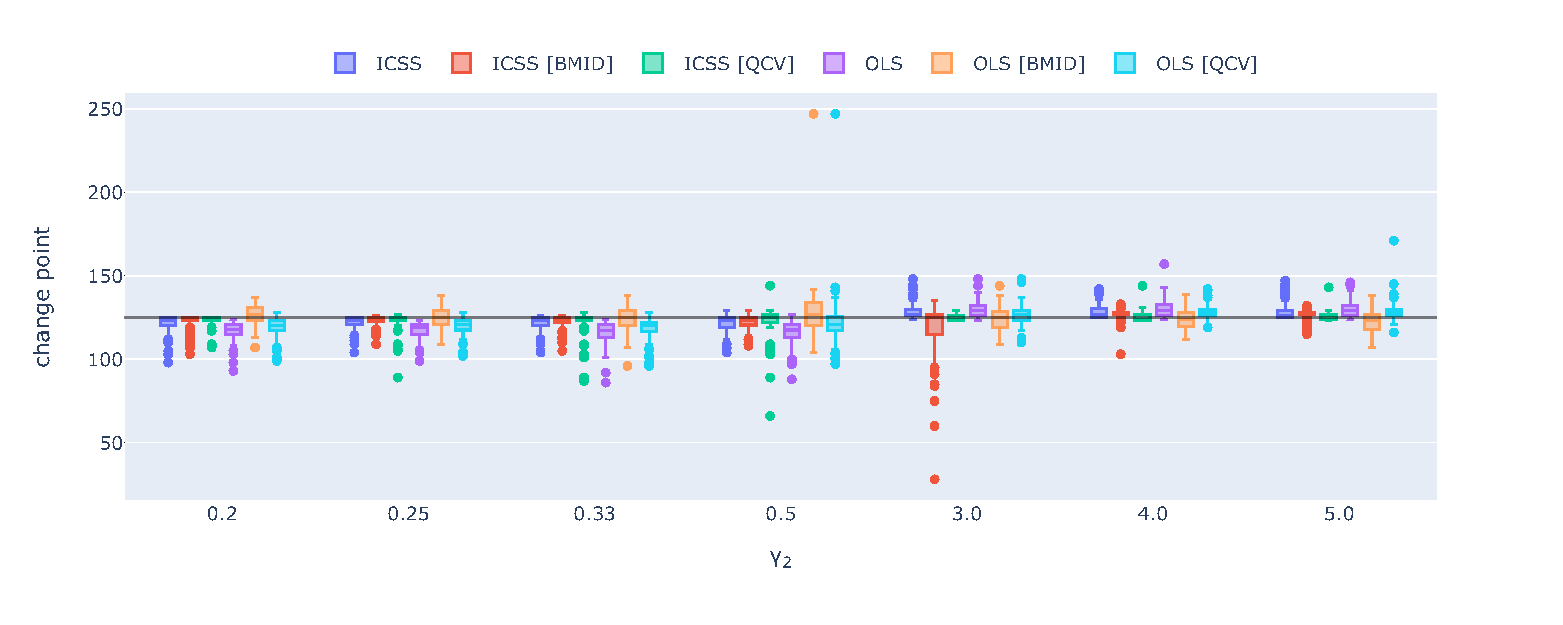}
    \caption{Detected change points obtained using selected procedures for sample described by model \eqref{eq:formula_stable} with size $N=250$. Distribution of the data: symmetric $\alpha-$stable distribution with \uline{$\alpha=2$}. True change point is marked via grey line. }
    \label{fig:sim_as_alpha2_n250}
\end{figure}

\clearpage
\section{Mixture Gaussian model}

\begin{figure}[h!]
    \centering
    \includegraphics[width=0.95\textwidth]{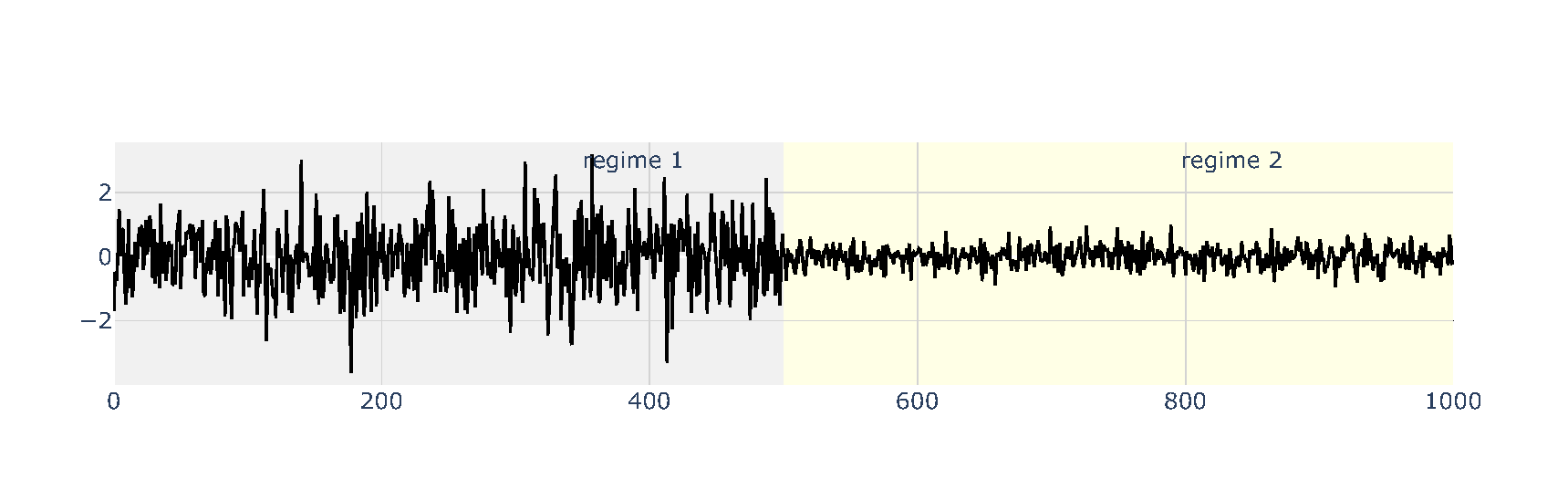}
    \caption{Example trajectory of data described by model \eqref{eq:mix_gauss_def} for $\nu=0.99, p=0.01, \gamma_1=1, \gamma_2=0.33.$}
    \label{fig:trajectory_mix_1}
\end{figure}

\begin{figure}[h!]
    \centering
    \includegraphics[width=0.95\textwidth]{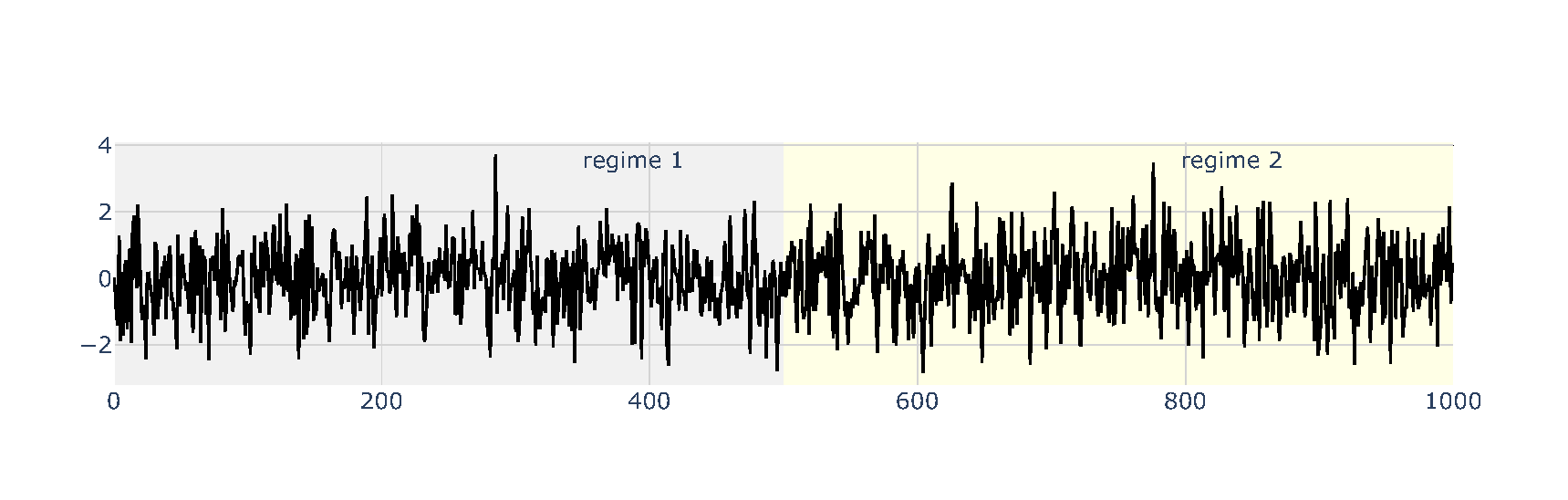}
    \caption{Example trajectory of data described by model \eqref{eq:mix_gauss_def} for $\nu=1.1, p=0.01, \gamma_1=1, \gamma_2=1.1.$}
    \label{fig:trajectory_mix_2}
\end{figure}

\vspace{-0.5cm}

\begin{figure}[h!]
    \centering
    \includegraphics[width=0.95\textwidth]{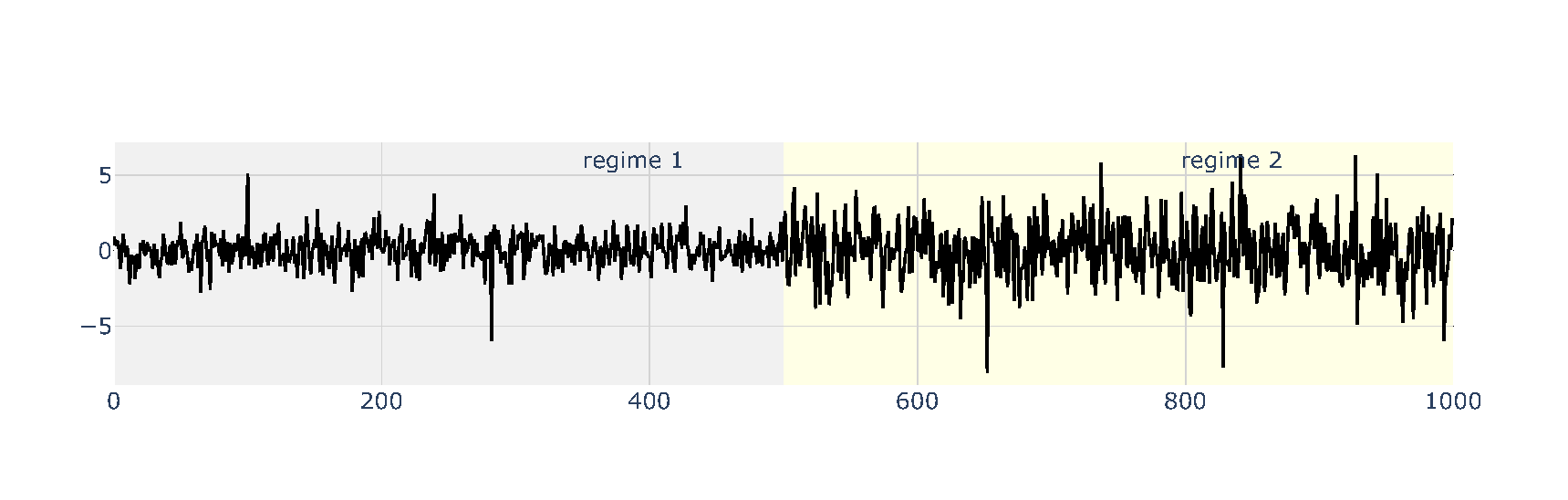}
    \caption{Example trajectory of data described by model \eqref{eq:mix_gauss_def} for $\nu=10, p=0.01, \gamma_1=1, \gamma_2=2.$}
    \label{fig:trajectory_mix_3}
\end{figure}

\vspace{-0.5cm}

\begin{figure}[h!]
    \centering
    \includegraphics[width=0.95\textwidth]{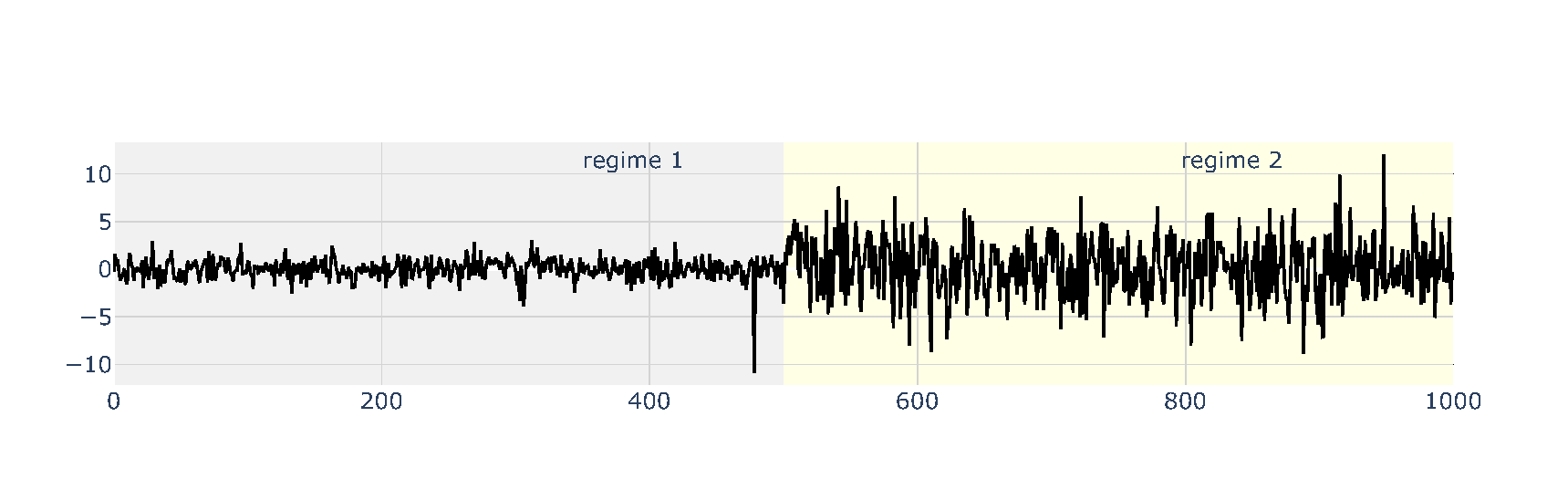}
    \caption{Example trajectory of data described by model \eqref{eq:mix_gauss_def} for $\nu=15, p=0.01, \gamma_1=1, \gamma_2=3.$}
    \label{fig:trajectory_mix_4}
\end{figure}

\begin{figure}[h!]
    \centering
    \includegraphics[width=0.8\textwidth]{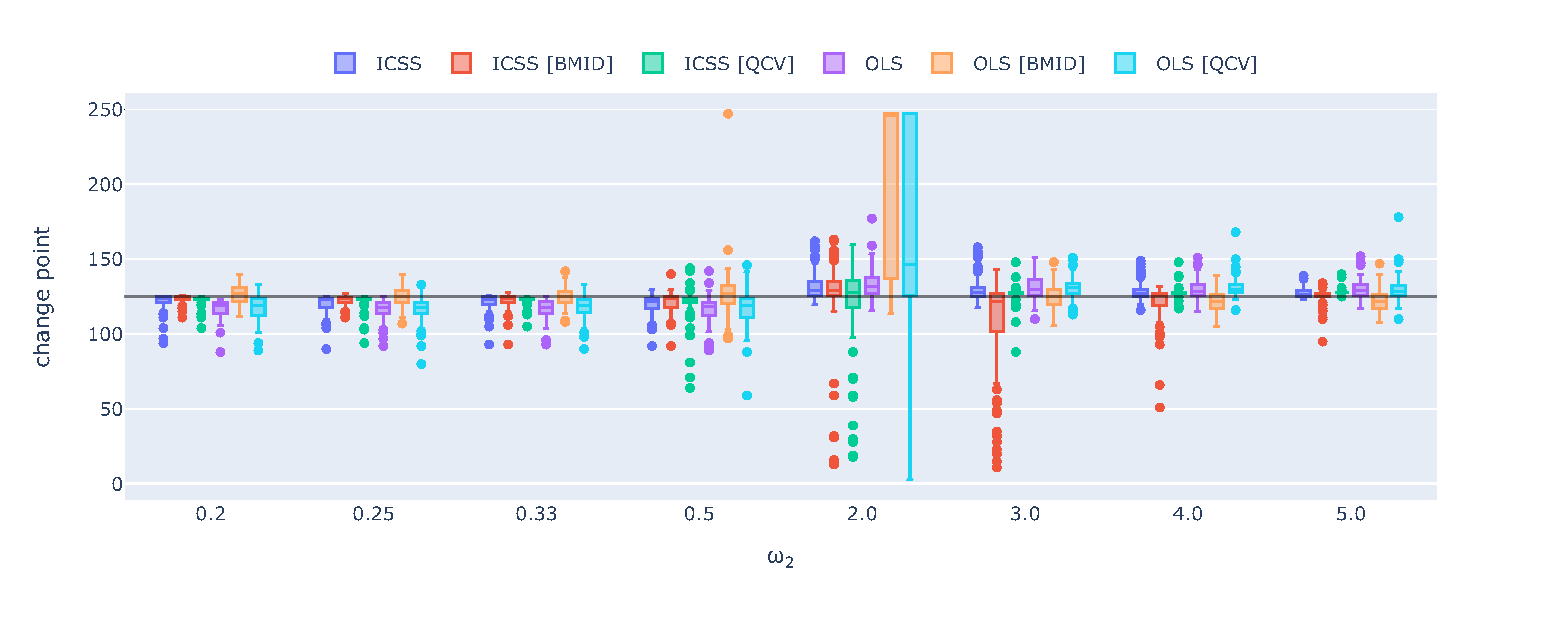}
    \caption{Detected change points obtained using selected procedures for sample described by model \eqref{eq:mix_gauss_def} with size $N=250$. Distribution of the data: the mixtured Gaussian model. Selected parameters: $p=0.05, \nu=1.5.$ True change point is marked via grey line. }
    \label{fig:sim_mixtured_gauss_v1_5_n250}
\end{figure}

\begin{figure}[h!]
    \centering
    \includegraphics[width=0.8\textwidth]{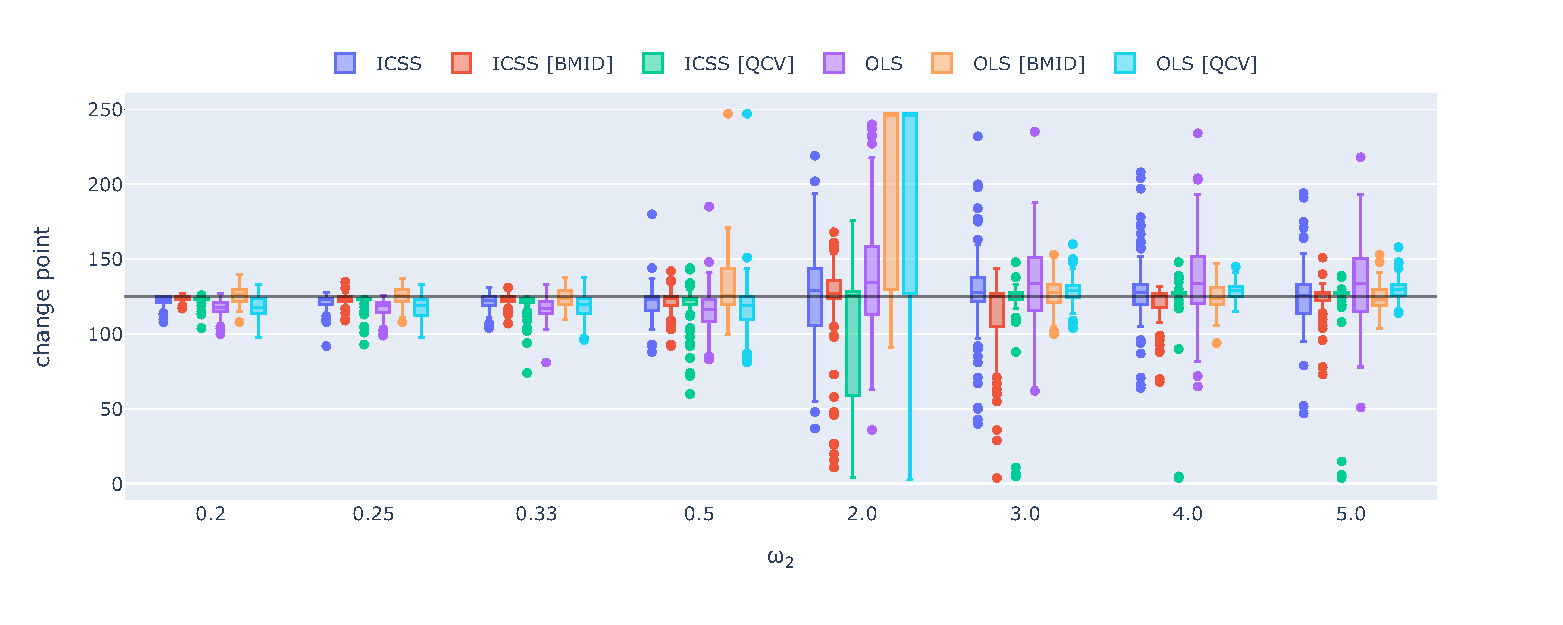}
    \caption{Detected change points obtained using selected procedures for sample described by model \eqref{eq:mix_gauss_def} with size $N=250$. Distribution of the data: the mixtured Gaussian model. Selected parameters: $p=0.05, \nu=5.$ True change point is marked via grey line. }
    \label{fig:sim_mixtured_gauss_v5_n250}
\end{figure}

\clearpage
\section{Time complexity analysis}
 
\begin{figure}[h!]
    \centering
    \includegraphics[width=0.95\textwidth]{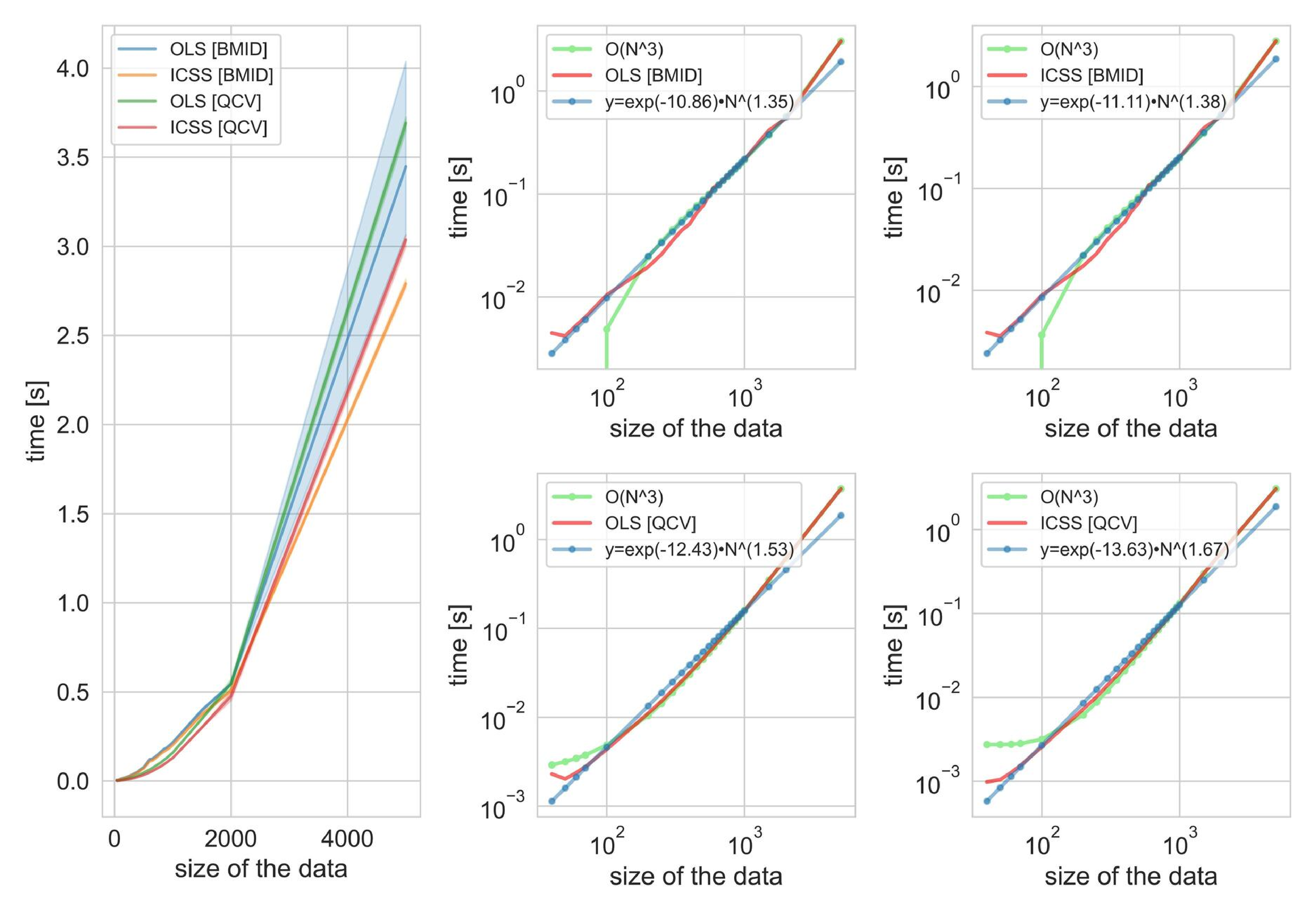}
    \caption{Analysis of computational time of the proposed procedures for data from model \eqref{eq:formula_stable}.}
    \label{fig:complexity_alpha_stable}
\end{figure}

\clearpage
\section{The mean value of MAE for simulated data}
\vspace{-0.5cm}
\begin{table}[h!]
\centering
\caption{MAE corresponding to the symmetric $\alpha-$stable case discussed in Section \ref{sec:alpha_stable_distr}. The MAE values are calculated based on $100$ Monte Carlo simulations of the samples of length  $N=1000$.}\label{tab_new1}
\begin{tabular}{clrrrrrr}
\hline
\multicolumn{1}{l}{$\alpha$} & $\gamma_2$ & ICSS   & ICSS\_BMID & ICSS\_QCV & OLS    & OLS\_BMID & OLS\_QCV \\ \hline
\multirow{8}{*}{1.1}         & 0.20       & 188.45 & 18.02      & 133.66    & 212.81 & 25.21     & 66.33    \\
                             & 0.25       & 184.04 & 18.84      & 109.19    & 210.83 & 34.07     & 65.59    \\
                             & 0.33       & 222.15 & 12.79      & 88.27     & 226.91 & 36.70     & 73.18    \\
                             & 0.50       & 226.03 & 44.42      & 121.09    & 233.40 & 51.29     & 92.90    \\
                             & 2.00       & 206.69 & 55.71      & 135.74    & 208.98 & 36.91     & 59.80    \\
                             & 3.00       & 209.72 & 10.87      & 143.65    & 221.29 & 38.27     & 55.57    \\
                             & 4.00       & 195.60 & 16.21      & 115.96    & 218.90 & 30.63     & 62.11    \\
                             & 5.00       & 192.84 & 11.29      & 98.12     & 205.09 & 33.29     & 84.42    \\ \hline
\multirow{8}{*}{1.9}         & 0.20       & 18.06  & 2.11       & 9.54      & 43.26  & 9.27      & 16.91    \\
                             & 0.25       & 15.85  & 2.15       & 13.70     & 38.74  & 8.04      & 15.20    \\
                             & 0.33       & 26.31  & 3.98       & 11.86     & 58.98  & 11.88     & 15.09    \\
                             & 0.50       & 55.37  & 6.74       & 17.77     & 74.76  & 15.66     & 16.06    \\
                             & 2.00       & 34.81  & 8.35       & 25.05     & 60.31  & 21.06     & 25.95    \\
                             & 3.00       & 24.46  & 4.25       & 10.40     & 54.17  & 9.67      & 13.10    \\
                             & 4.00       & 30.33  & 2.97       & 3.93      & 55.49  & 8.55      & 13.85    \\
                             & 5.00       & 29.72  & 2.91       & 7.72      & 48.51  & 11.37     & 11.96    \\ \hline
\multirow{8}{*}{2.0}         & 0.20       & 3.55   & 2.17       & 3.60      & 11.47  & 13.62     & 15.75    \\
                             & 0.25       & 3.13   & 2.12       & 3.64      & 14.49  & 10.81     & 17.58    \\
                             & 0.33       & 4.51   & 3.56       & 4.38      & 14.77  & 9.86      & 15.18    \\
                             & 0.50       & 5.35   & 4.91       & 7.38      & 14.91  & 11.56     & 17.72    \\
                             & 2.00       & 5.35   & 14.13      & 16.50     & 13.68  & 22.65     & 27.97    \\
                             & 3.00       & 3.37   & 5.99       & 5.05      & 9.58   & 8.59      & 11.89    \\
                             & 4.00       & 3.77   & 2.62       & 3.65      & 9.08   & 10.29     & 11.27    \\
                             & 5.00       & 3.82   & 2.29       & 3.48      & 9.28   & 16.20     & 12.08    \\ \hline
\end{tabular}
\end{table}

\begin{table}[h!]
\centering
\caption{MAE corresponding to the mixture Gaussian model case discussed in Section \ref{sec:mix_gauss_distr}. The MAE values are calculated based on $100$ Monte Carlo simulations of the samples of length  $N=1000$.}\label{tab_new2}
\begin{tabular}{clrrrrrr}
\hline
\multicolumn{1}{l}{$\nu / \omega_2$} & $\omega_2$ & ICSS  & ICSS\_BMID & ICSS\_QCV & OLS   & OLS\_BMID & OLS\_QCV \\ \hline
\multirow{8}{*}{1.5}                  & 0.20       & 3.53  & 1.55       & 3.48      & 12.94 & 15.01     & 17.92    \\
                                      & 0.25       & 4.04  & 2.97       & 4.36      & 13.76 & 10.65     & 16.75    \\
                                      & 0.33       & 4.02  & 3.20       & 4.59      & 12.95 & 8.80      & 18.80    \\
                                      & 0.50       & 7.60  & 6.46       & 7.50      & 17.39 & 14.16     & 22.34    \\
                                      & 2.00       & 7.11  & 11.02      & 17.90     & 14.47 & 22.78     & 23.89    \\
                                      & 3.00       & 4.35  & 13.73      & 2.89      & 13.30 & 9.33      & 17.52    \\
                                      & 4.00       & 4.23  & 4.25       & 3.66      & 11.10 & 10.96     & 15.55    \\
                                      & 5.00       & 3.68  & 3.37       & 3.67      & 10.04 & 16.47     & 16.61    \\ \hline
\multirow{8}{*}{5.0}                  & 0.20       & 3.94  & 1.69       & 2.81      & 12.16 & 13.85     & 16.08    \\
                                      & 0.25       & 4.13  & 2.18       & 3.61      & 13.65 & 9.82      & 17.80    \\
                                      & 0.33       & 4.72  & 3.47       & 5.02      & 13.71 & 10.14     & 19.04    \\
                                      & 0.50       & 7.29  & 5.91       & 6.98      & 20.80 & 14.70     & 23.69    \\
                                      & 2.00       & 23.03 & 15.56      & 16.00     & 40.69 & 21.35     & 27.89    \\
                                      & 3.00       & 15.42 & 12.25      & 4.49      & 33.39 & 11.19     & 17.15    \\
                                      & 4.00       & 15.29 & 6.93       & 3.65      & 29.70 & 11.67     & 17.08    \\
                                      & 5.00       & 11.52 & 6.10       & 3.59      & 28.44 & 14.39     & 16.05    \\ \hline
\end{tabular}
\end{table}

\end{document}